\documentclass[journal=jacsat,manuscript=article]{achemso}

\usepackage[version=3]{mhchem} 
\usepackage{multirow}
\usepackage{xcolor}
\usepackage{indentfirst}
\usepackage[T1]{fontenc}     
\usepackage{lmodern}         
\usepackage{amsmath}
\usepackage{booktabs}
 \usepackage{hyperref}

\author{José A. S. Laranjeira}
\affiliation[Unesp]
{Modeling and Molecular Simulation Group, São Paulo State University (UNESP), School of Sciences, Bauru, 17033-360, SP, Brazil}
\email{jose.laranjeira@unesp.br}
\author{Nicolas F. Martins}
\affiliation[UNESP]
{Modeling and Molecular Simulation Group, São Paulo State University (UNESP), School of Sciences, Bauru, 17033-360, SP, Brazil}
\author{Kleuton A. L. Lima}
\affiliation[Campinas]
{Department of Applied Physics and Center for Computational Engineering and Sciences, State University of Campinas, Campinas, 13083-859, SP, Brazil}
\author{Luis A. Cabral}
\affiliation[UNESP1]
{Department of Physics and Meteorology, São Paulo State University (UNESP), School of Sciences, Bauru, 17033-360, SP, Brazil}
\author{Luiz A. Ribeiro}
\affiliation[IF]{Institute of Physics, University of Bras{\'{i}}lia, 70919-970, Bras{\'{i}}lia, DF, Brazil.}
\alsoaffiliation[lccmat]{Computational Materials Laboratory, LCCMat, Institute of Physics, University of Bras{\'{i}}lia, 70919-970, Bras{\'{i}}lia, DF, Brazil.}
\author{Douglas S. Galvão}
\affiliation[Campinas]
{Department of Applied Physics and Center for Computational Engineering and Sciences, State University of Campinas, Campinas, 13083-859, SP, Brazil}
\author{Julio R. Sambrano}
\affiliation[UNESP]
{Modeling and Molecular Simulation Group, São Paulo State University (UNESP), School of Sciences, Bauru, 17033-360, SP, Brazil}
\email{jr.sambrano@unesp.br}

\title[An \textsf{achemso} demo]
  {TPHE-Graphene: A First-Principles Study of a New 2D Carbon Allotrope for Hydrogen Storage}


\begin{document}

\begin{abstract}
The shift from fossil fuels to renewable energy sources is essential for reducing global carbon emissions and addressing climate change. Developing advanced materials for efficient hydrogen storage enables sustainable energy solutions in this context. Herein, we propose sodium-decorated TPHE-graphene as a high-performance two-dimensional material for hydrogen storage. Density functional theory (DFT) calculations demonstrate that TPHE-graphene exhibits dynamical, thermal, energetic, and mechanical stability, as confirmed by cohesive energy, phonon dispersion, and molecular dynamics simulations. The monolayer displays metallic behavior and a high Young's modulus of 250.46 N/m. Upon sodium decoration, strong chemisorption occurs with a binding energy of -2.08 eV and minimal tendency for Na atom clustering. Hydrogen adsorption analysis reveals that each Na atom can bind up to five H$_2$ molecules, resulting in a gravimetric storage capacity of 9.52 wt\%. The calculated H$_2$ adsorption energies range from -0.22 eV to -0.18 eV, falling within the ideal range for reversible adsorption under ambient conditions. These findings highlight Na-decorated TPHE-graphene as a structurally robust and efficient hydrogen storage material well-suited for future green energy applications.
\end{abstract}

\section{Introduction}

Transitioning from traditional fossil fuels to renewable energy sources is essential for reducing global carbon emissions and mitigating associated environmental impacts and health issues \cite{pani2022importance}. In this context, hydrogen (H$_2$) stands out as an environmentally friendly energy carrier capable of meeting the demands of modern technologies \cite{oliveira2021green, laimon2024towards}. Unlike other clean energy sources such as solar and wind, hydrogen-based energy systems are not weather-dependent. However, a significant challenge lies in the efficient storage of H$_2$, given its low volumetric and gravimetric energy densities. Depending on the storage method, hydrogen often needs to be stored under high-pressure conditions \cite{ma2024large, dewangan2022comprehensive}.

Since the seminal report by Reilly and Wiswall \cite{reilly1968reaction} demonstrating the excellent hydrogen storage performance of Mg$_2$NiH$_4$, solid-state hydrogen storage has gained significant attention \cite{rusman2016review}. Despite continued exploration of various platforms such as nanotubes \cite{pinjari2023mechanism,cardoso2021lithium}, metal-organic frameworks (MOFs) \cite{shet2021review,kim2024comparing}, and metal hydrides \cite{bishnoi2024architectural}, many of these materials suffer from low gravimetric hydrogen capacity, primarily due to their high molecular weights. To address this limitation, researchers have proposed tuning the nanostructure dimensionality of host materials to improve hydrogen uptake \cite{schneemann2018nanostructured}.

Two-dimensional (2D) materials, in particular, have emerged as promising candidates for sustainable energy applications, owing to their high surface-area-to-weight ratio and ultralight atomic structures \cite{abifarin20242d, qin2023two,lokesh2022advanced}. Graphene-like materials exemplify these characteristics. However, the intrinsically low reactivity of pristine carbon networks leads to weak interactions with hydrogen molecules, resulting in limited adsorption efficiency \cite{zhang2012ultra}.

A common strategy to improve the interaction between hydrogen molecules and two-dimensional (2D) carbon-based materials is metal decoration. In particular, alkali earth metals (Be, Mg, Ca) and alkali metals (Li, Na, K) offer a promising approach due to their low atomic masses, which align with the criteria for high gravimetric hydrogen capacity \cite{gopalsamy2014hydrogen, mohajeri2018light}. Moreover, metal decoration typically induces a physisorption regime characterized by optimal adsorption energies of --0.10 to --0.40 eV \cite{yadav2016study}, suitable for reversible hydrogen uptake under ambient conditions.

During the past decade, considerable effort has been dedicated to exploring novel 2D carbon allotropes for hydrogen storage applications \cite{yadav2015first, li2011high, srinivasu2012graphyne}. The recently synthesized biphenylene sheet has demonstrated excellent performance in retaining H$_2$ molecules \cite{kaewmaraya2023ultrahigh, ma2024li}. First-principles calculations have further confirmed the high hydrogen adsorption capacities in various functionalized 2D carbon materials, including irida-graphene \cite{zhang2024decorated, zhang2024li}, PAI-graphene \cite{mahamiya2023potential}, PHE-graphene \cite{LARANJEIRA2025139}, holey-graphyne \cite{adithya2024decorated} and M-graphene \cite{chen2025penta}.

Recently, Shi \textit{et al.} \cite{doi:10.1021/acs.jpclett.1c03193} employed a random structure search strategy combined with graph theory to explore novel two-dimensional (2D) $sp^2$ carbon allotropes. Their study identified 1114 previously unreported carbon structures, among which 190 were classified as Dirac semimetals, 241 as semiconductors, and 683 as conventional metals. These materials exhibit various exotic electronic behaviors, including type III, type I/II mixed, and type I/III mixed semimetallic characteristics.

From this extensive dataset, we focus on one particular structure, TPHE-graphene, among the configurations proposed in their work. TPHE-graphene presents a unique 2D lattice composed of square, pentagonal, hexagonal, and enneagonal carbon rings arranged in a rectangular geometry and belonging to the \textit{Pmma} space group (No. 51). Our particular interest is the spatial distribution of its enneagonal rings, which are well-separated and symmetrically positioned—creating an ideal topology for anchoring metal adatoms and, consequently, enhancing hydrogen (H$_2$) adsorption through decoration strategies.

In response to the growing demand for novel 2D materials that surpass the minimum requirements for hydrogen (H$_2$) storage capacity, we propose sodium-decorated TPHE-graphene as a promising candidate, based on first-principles calculations within the density functional theory (DFT) framework. Sodium (Na) was selected for decoration due to its low atomic mass, high reactivity, and low melting point, contributing to enhanced interactions with H$_2$ molecules.

Herein, we systematically investigate the hydrogen adsorption behavior of Na-decorated TPHE-graphene through a comprehensive set of computational analyses, including molecular dynamics simulations, charge density difference (CDD), projected density of states (PDOS), adsorption and consecutive adsorption energy calculations, hydrogen adsorption capacity (HAC), and thermodynamic modeling. Furthermore, we assessed the dynamical, thermal, and mechanical stability of the pristine TPHE-graphene monolayer, providing valuable insights into its viability as a high-performance hydrogen storage substrate.

\section{Methodology}

First-principles calculations based on DFT were performed to investigate the structural and hydrogen storage properties of TPHE-graphene. Exchange-correlation effects were treated using the generalized gradient approximation (GGA) as proposed by Perdew, Burke, and Ernzerhof (PBE)~\cite{PhysRevLett.77.3865,ernzerhof1999assessment}. The projector-augmented wave (PAW) method~\cite{PhysRevB.50.17953} was used for the interaction between the core and the valence electrons. All simulations were performed using the Vienna \textit{ab initio} Simulation Package (VASP)~\cite{kresse1993ab,kresse1996efficient}. The set of plane-wave basis was defined with an energy cut-off of 520~eV. To avoid spurious interactions due to periodic boundary conditions, a vacuum layer of 15~\r{A} was added along the $z$-direction.

Structural optimization and projected density of states (PDOS) calculations were carried out using $6 \times 6 \times 1$ and $9 \times 9 \times 1$ \textbf{k}-point meshes centered at the $\Gamma$-point, respectively. The van der Waals interactions were taken into account using the Grimme DFT-D2 method~\cite{grimme2006semiempirical}. Geometry optimization used the conjugate gradient algorithm with convergence thresholds of $10^{-5}$~eV for total energy and 0.01~eV/\r{A} for atomic forces. Charge transfer analysis was conducted using the Bader partitioning scheme.

Molecular dynamics (MD) simulations were performed using the tight-binding approximation available in the DFTB+ package \cite{hourahine2020dftb+}, employing the 3ob parametrization \cite{doi:10.1021/ct300849w} and the D4 dispersion method \cite{caldeweyher2017extension}. These MD simulations were performed with the Berendsen thermostat \cite{berendsen1984molecular} at 300 K for 5 ps and a time step of 1 fs.

The energetic stability of TPHE-graphene was analyzed by its cohesive energy (\(E_{\text{coh}}\)):

\begin{equation}
E_{\text{coh}} = \frac{E_{\text{TPHE}} - \sum_i n_iE_i}{\sum_i n_i},
\end{equation}

\noindent In this formulation, \(E_{\text{TPHE}}\) stands for the total energy of the system, while \(E_i\) represents the energy of an isolated carbon atom, and \(n_i\) denotes the number of C atoms in the structure. The same procedure was consistently applied to evaluate the cohesive energy (\(E_{\text{coh}}\)) of the other 2D materials examined throughout this study.

The charge density difference (CDD) obtained from the Na@TPHE-graphene system was calculated by:

\begin{equation}
\Delta \rho = \rho_{\text{(Na@TPHE})} - \rho_{\text{(Na)}} - \rho_{\text{(TPHE)}},
\label{eq:charge_density}
\end{equation}

\noindent where $\rho_{(\text{Na@TPHE})}$, $\rho_{(\text{Na})}$, and $\rho_{(\text{TPHE})}$ refer to the charge densities of the Na@TPHE-graphene substrate, isolated Na adatoms, and pristine TPHE-graphene monolayer, respectively. For the hydrogen-adsorbed Na@TPHE-graphene system, the CDD was obtained as:

\begin{equation}
\Delta \rho = \rho_{(\text{Na@TPHE + H$_2$})} - \rho_{(\text{H$_2$})} - \rho_{(\text{Na@TPHE})},
\label{eq:charge_density1}
\end{equation}

\noindent where $\rho_{(\text{Na@TPHE + H$_2$})}$, $\rho_{(\text{H$_2$})}$, and $\rho_{(\text{Na@TPHE})}$ refer to the charge densities of the Na@TPHE-graphene with H$_2$ molecules, isolated H$_2$ molecules, and Na@TPHE-graphene substrate, respectively.

The adsorption energy ($E_{\text{ads}}$) for Na@TPHE-graphene + $n \textrm{H}_2$ systems was computed using the following expression:

\begin{equation}
\begin{split}
E_{\text{ads}} &= \\\frac{1}{n} &\Big( E_{\text{Na@TPHE + $n \textrm{H}_2$}}
\quad - E_{\text{Na@TPHE}} - n \textrm{H}_2 \Big),
\end{split}
\end{equation}

\noindent where $E_{\text{Na@TPHE + $n \textrm{H}_2$}}$ represents the total energy of the Na@TPHE-graphene system with $n$ adsorbed H$_2$ molecules, $E_{\text{Na@TPHE}}$ is the energy of the bare Na@TPHE-graphene substrate, and $E_{\text{H}_2}$ corresponds to the energy of an isolated H$_2$ molecule.

The consecutive adsorption energy ($E_{\text{con}}$) for each H$_2$ molecules addition on the surface was determined as:

\begin{equation}
\begin{split}
E_{\text{con}}& =\\&\hspace{-0.5cm}\frac14\left(E_{\text{Na@TPHE} + n \textrm{H}_2}\right) - 
E_{\text{Na@TPHE} + (n-4) \textrm{H}_2}  - 4E_{\text{H}_2}.
\end{split}
\end{equation}

The hydrogen adsorption capacity (HAC) in weight percentage was calculated as:

\begin{equation}
\text{HAC} (\text{wt\%}) = \frac{ n_\text{H} M_\text{H}}{n_\text{C} M_\text{C} + 
n_\text{Na} M_\text{Na} + n_\text{H} M_\text{H} },
\end{equation}

\noindent where $n_X$ and $M_X$ represent the number of atoms and molar masses of element $X$ ($X = \text{H}, \text{C}, \text{Na}$), respectively.

\begin{figure*}[!ht]
    \centering
    \includegraphics[width=0.8\linewidth]{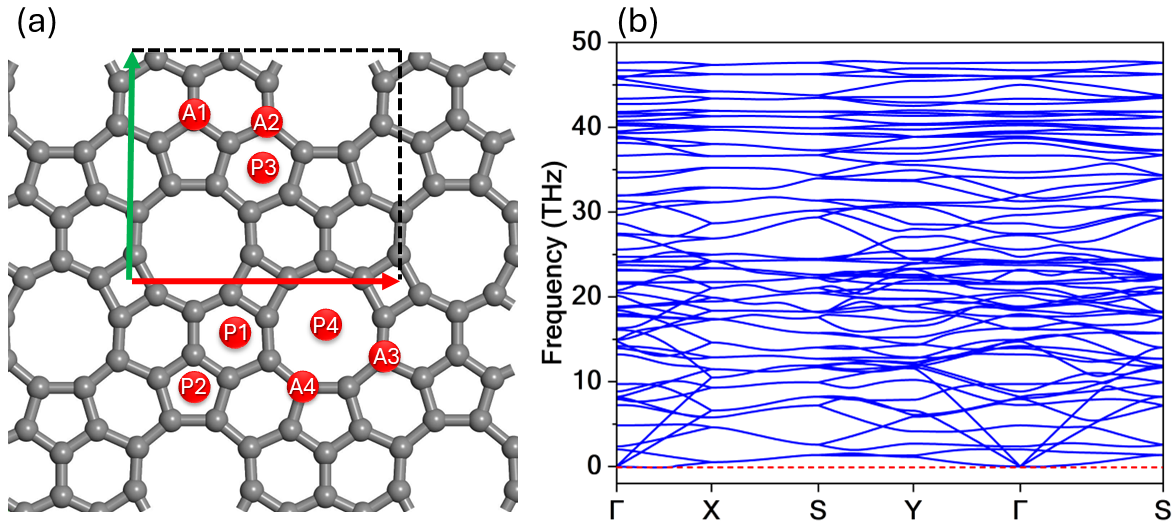}
    \caption{(a) Top view of the TPHE-graphene monolayer highlighting the rectangular unit cell (dashed black lines) and the evaluated high-symmetry adsorption sites for Na decoration. Sites labeled P1--P4 correspond to pore-centered positions, while A1--A4 indicate atomic sites. (b) Phonon dispersion of pristine TPHE-graphene.}
    \label{fig:1}
\end{figure*}

Assuming atmospheric pressure (1 atm), the hydrogen desorption temperature ($T_{\text{R}}$) was estimated using the van't Hoff equation \cite{durbin2013review, alhameedi2019metal}:

\begin{equation}
T_{\text{des}} = \left| E_{\text{ads}} \right|\frac{R}{K_B \Delta S},
\end{equation}

\noindent where $R$ denotes the universal gas constant, $k_B$ represents the Boltzmann constant, and $\Delta S$ corresponds to the change in entropy during the hydrogen phase transition from gas to liquid, taken as 75.44~J~mol$^{-1}$~K$^{-1}$.

A thermodynamic analysis was performed to evaluate the adsorption and desorption behavior of H$_2$ molecules under realistic conditions, employing the grand canonical partition function $Z$, given by:

\begin{equation}
   Z = 1 + \sum_{i=1}^{n} \exp\left(-\frac{E_i^{\text{ads}} - \mu}{k_B T}\right).
\end{equation}

In this equation, $n$ denotes the total number of H$_2$ molecules that can be adsorbed, while $\mu$ represents the chemical potential of a hydrogen molecule in the gaseous state. Here, $E_{i}^{\text{ads}}$ corresponds to the adsorption energy of the $i$-th H$_2$ molecule \cite{hashmi2017ultra, kaewmaraya2023ultrahigh}.

\section{Results}

\subsection{Structural, electronic and mechanical features of TPHE-graphene}

As discussed previously, TPHE-graphene features a rectangular lattice that crystallizes in the \textit{Pmma} space group (No. 51). Its atomic structure comprises an arrangement of square, pentagonal, hexagonal, and enneagonal carbon rings, as illustrated in Fig.~\ref{fig:1}(a). The optimized lattice parameters were calculated as $\vec{a} = 8.75$~\r{A} and $\vec{b} = 7.79$~\r{A}, and the unit cell contains nine non-equivalent carbon atoms.

To evaluate the energetic stability of TPHE-graphene, we computed its cohesive energy ($E_\text{coh}$), obtaining a value of -7.63~eV/atom. This result is comparable to the cohesive energies of other known 2D carbon allotropes, such as graphene (–7.68~eV/atom), PHE-graphene (–7.56~eV/atom), T-graphene (–7.45~eV/atom), and penta-graphene (–7.13~eV/atom). Based on this comparison with ground-state atomic energies, TPHE-graphene can be classified as energetically stable.

We now examine the phonon dispersion of TPHE-graphene, as presented in Fig.~\ref{fig:1}(b). The absence of negative frequencies (i.e., imaginary modes) confirms the dynamical stability of the monolayer. Moreover, the presence of multiple phonon band crossings suggests the existence of several thermal transport channels, which may contribute to enhanced thermal conductivity. The phonon spectrum extends up to 50~THz, a typical range for carbon-based materials with $sp^2$ hybridization.

\begin{figure*}
    \centering
    \includegraphics[width=0.8\linewidth]{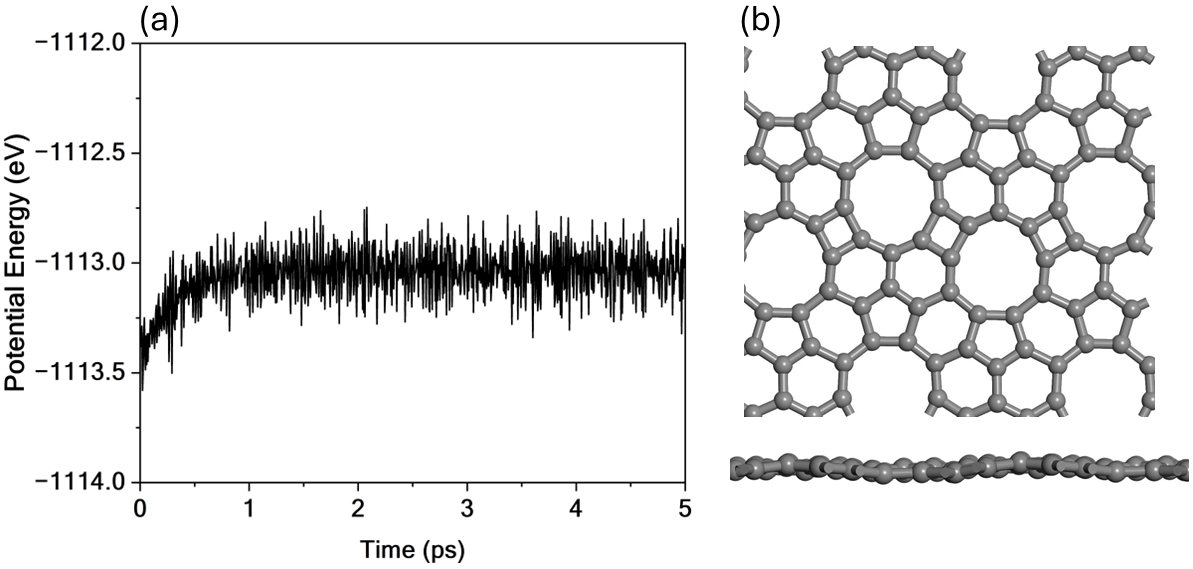}
    \caption{MD simulation results for pristine TPHE-graphene at 300~K. (a) Time evolution of the potential energy over a 5 ps simulation. (b) Top and side views of the final structure.}
    \label{fig:md_pristine}
\end{figure*}

MD simulations were performed to further assess the thermal stability of TPHE-graphene. Figure~\ref{fig:md_pristine} shows the potential energy profile over time (\ref{fig:md_pristine}(a)) and the final structure after 5~ps of simulation at 300~K (\ref{fig:md_pristine}(b)). The potential energy stabilizes after approximately 0.5~ps, oscillating around a mean value of -1113.2~eV, with fluctuations not exceeding 0.5~eV, as illustrated in Fig. \ref{fig:md_pristine}(a). This stability indicates the absence of reconstruction or phase transition events throughout the simulation. Additionally, the final geometry confirms that the structural integrity of the monolayer is preserved, with only minimal distortions observed, most notably a slight buckling in the lattice  (see Fig. \ref{fig:md_pristine}(b)).  

\begin{figure*}
    \centering
    \includegraphics[width=0.8\linewidth]{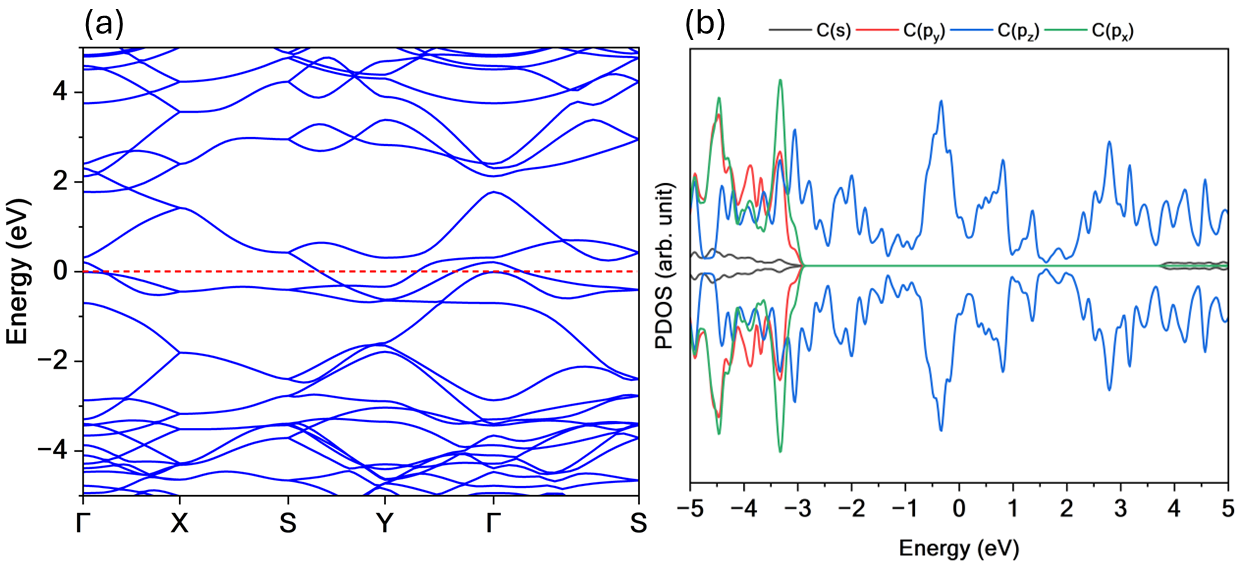}
    \caption{(a) Band structure and (b) PDOS for TPHE-graphene system. This novel monolayer exhibits metallic behavior, characterized by several bands that cross the Fermi level (red dashed line).}
    \label{fig:band+dos}
\end{figure*}

The electronic structure of TPHE-graphene was investigated through its band structure and PDOS, as shown in Fig.~\ref{fig:band+dos}. The results reveal a metallic character, evidenced by three bands that cross the Fermi level. The highest fully occupied bands exhibit significant dispersion (approximately 2~eV), correlating with a noticeable reduction in the PDOS within that energy range. Near the Fermi level (indicated by the red dashed line), a concentration of partially occupied states accounts for the non-zero PDOS, further supporting the metallic nature of the system.

Additionally, two tilted cone-like crossings appear along the \( \Gamma \rightarrow X \) and \( S \rightarrow Y \) directions. Although these features do not lie exactly at the Fermi level, they originate from partially filled bands and are of particular interest due to their potential to host anisotropic charge carriers and support unconventional transport phenomena. In the conduction band, a strong band dispersion is also observed near the Fermi level, accompanied by a corresponding dip in the PDOS, reinforcing the delocalized nature of the electronic states in this region.

The orbital composition of the electronic structure was analyzed using PDOS. For lower energies in valence band, the electronic states are primarily composed of C($p_x$) and C($p_y$) orbitals, with a minor contribution from C($s$) orbitals, indicating the formation of localized $\sigma$ bonds. For energies above -3~eV, the PDOS is dominated by C($p_z$) orbitals, which are associated with the delocalized $\pi$ system—characteristic of $sp^2$-hybridized carbon frameworks. This $\pi$-dominance near the Fermi level reinforces the metallic and conjugated nature of the TPHE-graphene monolayer.

\begin{figure*}
    \centering
    \includegraphics[width=1\linewidth]{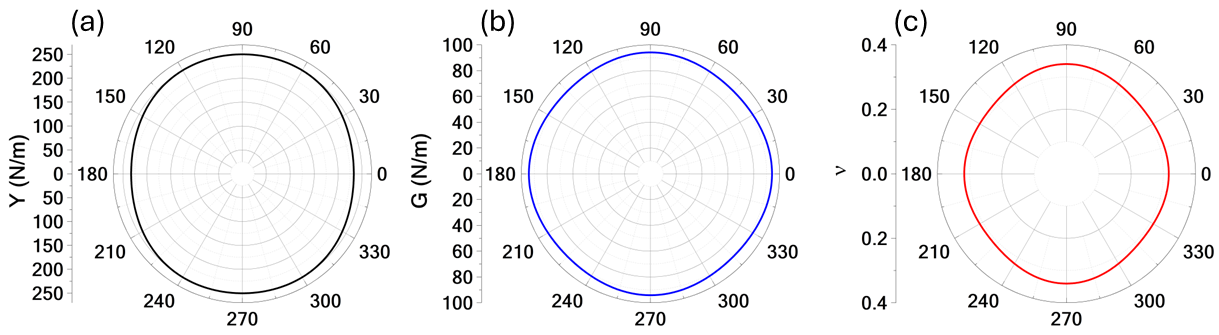}
    \caption{(a) Young's modulus (\( Y \)), (b) shear modulus (\( G \)), and (c) Poisson's ratio (\( \nu \)) for TPHE-graphene.}
    \label{fig:mech}
\end{figure*}

The mechanical properties of TPHE-graphene were evaluated and are illustrated in terms of the Young's modulus ($Y$), shear modulus ($G$), and Poisson's ratio ($\nu$) in Fig.~\ref{fig:mech}. The Young's modulus reaches a maximum (minimum) of 250.46~N/m (232.74~N/m), with an anisotropy ratio of 1.08. Similarly, the $G$ varies slightly between 94.12~N/m and 90.77~N/m, yielding an anisotropy ratio of 1.04. The Poisson's ratio exhibits minimal directional dependence, varying from 0.31 to 0.34, with an anisotropy ratio of 1.10. These slight variations indicate that TPHE-graphene behaves as an almost mechanically isotropic material.

Mechanical stability was further assessed by computing the elastic constants, resulting in $C_{11} = 260.87$~N/m, $C_{22} = 280.32$~N/m, $C_{12} = 88.80$~N/m, and $C_{66} = 94.13$~N/m. For a 2D material with rectangular symmetry, the Born–Huang stability criteria \cite{PhysRevB.90.224104,doi:10.1021/acs.jpcc.9b09593} require that $C_{11} > 0$, $C_{66} > 0$, and $C_{11} C_{22} > C_{12}^2$—all of which are satisfied by TPHE-graphene, confirming its mechanical robustness.

\subsection{Sodium decoration on TPHE-graphene monolayer}

\begin{figure}
    \centering
    \includegraphics[width=0.6\linewidth]{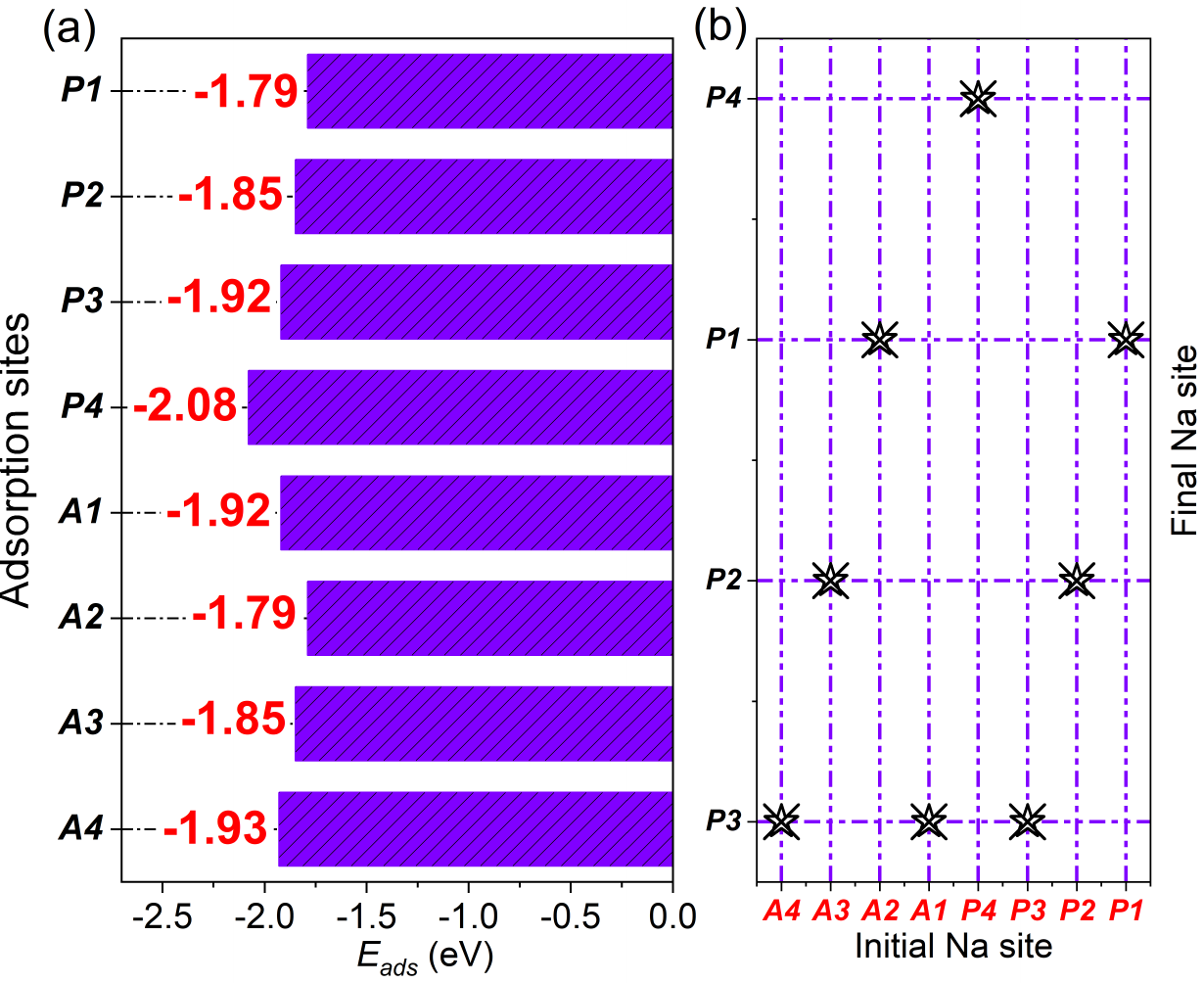}
    \caption{Adsorption energies (E$_\text{ads}$) for each adsorption site along with final configurations for the adsorption sites evaluated during Na decoration on the TPHE-graphene. Strong adsorption is also observed at the other pore sites, with $E_{\text{ads}}$ values of $-1.79$~eV/atom, $-1.85$~eV/atom, and $-1.92$~eV/atom for the P1, P2, and P3 sites, respectively.}
    \label{fig:single_ads}
\end{figure}

In this section, we investigate the sodium decoration mechanism on the TPHE-graphene monolayer. Specifically, we evaluated the adsorption of Na atoms at eight high-symmetry sites: P1, P2, P3, and P4, associated with the pores (ring centers) of the lattice, and A1, A2, A3, and A4, which correspond to on-top atomic positions within the carbon framework (refer to Fig. \ref{fig:1}). The computed adsorption energies and final configurations are summarized in Table~\ref{tbl2} and visually represented in Fig.~\ref{fig:single_ads}.

\begin{figure}
    \centering
    \includegraphics[width=0.4\linewidth]{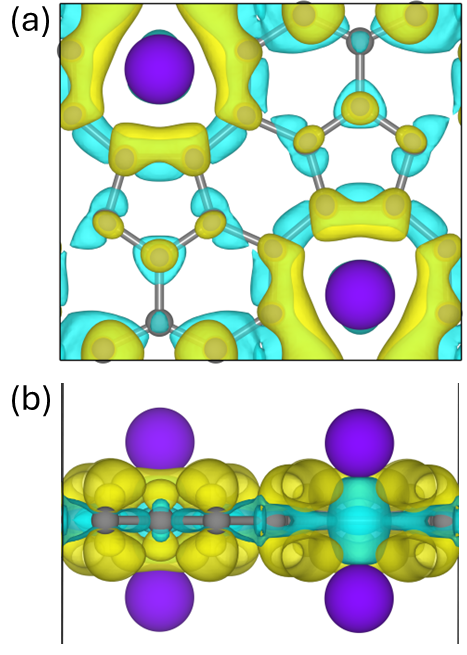}
    \caption{Charge density difference (CDD) visualizations shown from the (a) top and (b) side views for Na@TPHE-graphene system. The yellow (blue) regions indicate charge accumulation (depletion).}
    \label{fig:CDD_a@TPHE-graphene}
\end{figure}

Our results indicate that Na adatoms consistently favor adsorption at the pore-centered sites of the TPHE-graphene lattice, irrespective of their initial positions. Among these, the enneagonal pore (P4) emerges as the most favorable adsorption site, exhibiting an adsorption energy ($E_{\text{ads}}$) of -2.08~eV/atom. This value reflects a strong chemisorption interaction, which promotes the formation of a stable Na@TPHE-graphene complex when the monolayer is fully decorated at the P4 sites. As illustrated in Fig.~\ref{fig:CDD_a@TPHE-graphene}, this configuration comprises four Na atoms per TPHE-graphene unit cell.

Strong adsorption is also observed at the remaining pore sites, with $E_{\text{ads}}$ values of -1.79~eV/atom, -1.85~eV/atom, and -1.92~eV/atom for the P1, P2, and P3 sites, respectively. These relatively high adsorption energies indicate limited mobility of Na adatoms on the TPHE-graphene surface. Consequently, a high diffusion barrier is anticipated, effectively suppressing adatom clustering and contributing to the Na-decorated monolayer's structural stability.

\begin{table}[!ht]
\centering
\caption{Adsorption energies (E$_\text{ads}$) and final configurations for the adsorption sites evaluated during Na decoration on the TPHE-graphene.}
\label{tbl2}
\begin{tabular*}{\linewidth}{@{\extracolsep{\fill}} lcc}
\toprule
\textbf{Initial site} & \textbf{E$_\text{ads}$ (eV)} & \textbf{Final site} \\
\midrule
P1 &  -1.79  & P1  \\
P2 &  -1.85  & P2  \\
P3 &  -1.92  & P3  \\
P4 &  -2.08  & P4   \\
A1 &  -1.92  & P3  \\
A2 &  -1.79  & P1  \\
A3 &  -1.85  & P2  \\
A4 &  -1.93  & P3  \\
\bottomrule
\end{tabular*}
\label{tab:sites}
\end{table}

To further elucidate the nature of the interaction between Na adatoms and the TPHE-graphene monolayer, we analyzed the charge density difference (CDD) map for the Na@TPHE-graphene system, as depicted in Fig.~\ref{fig:CDD_a@TPHE-graphene}. In addition, a Bader charge analysis was conducted to quantify the charge transfer resulting from the adsorption process. The CDD map shows pronounced charge accumulation around the enneagonal pores where Na atoms are adsorbed, indicating a net electron transfer from the adatoms to the graphene surface. This observation is corroborated by the Bader analysis, which reveals a charge transfer of approximately $-0.75~|e|$ per Na atom to the monolayer. Such substantial charge transfer supports an ionic adsorption mechanism and accounts for the strong binding affinity between Na and the TPHE-graphene structure.

\begin{figure*}
    \centering
    \includegraphics[width=0.8\linewidth]{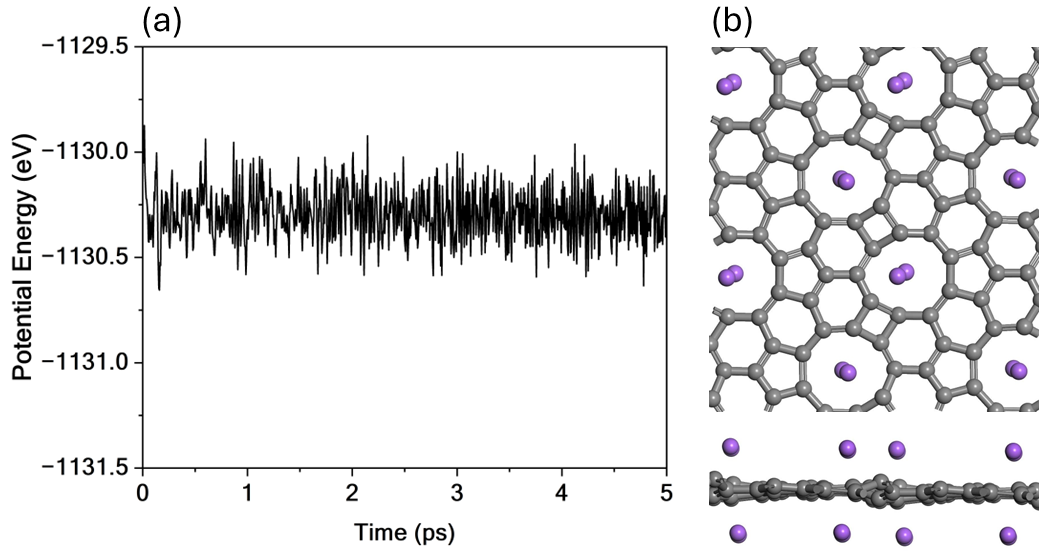}
    \caption{MD simulation results for pristine Na@TPHE-graphene at 300 K. (a) Time evolution of the potential energy and (b) final system configuration for Na@TPHE-graphene.}
    \label{fig:md_TPHE_NA}
\end{figure*}

A fundamental requirement for a material to serve as a viable hydrogen storage substrate is its thermal stability under ambient conditions. To assess the thermal stability of Na@TPHE-graphene, molecular dynamics (MD) simulations were conducted at 300K for 5ps, as illustrated in Fig.~\ref{fig:md_TPHE_NA}. In the simulation, the potential energy fluctuates around a stable mean value, with oscillation amplitudes confined to approximately 0.5~eV. This stability suggests that no desorption events or structural reconstructions occurred throughout the simulation. Furthermore, visual inspection of the final structure reveals that the Na atoms remain firmly anchored at their most favorable adsorption sites, exhibiting only minor positional deviations. These observations collectively confirm the thermal robustness of the Na-decorated TPHE-graphene system at room temperature.

\begin{figure}
    \centering
    \includegraphics[width=0.7\linewidth]{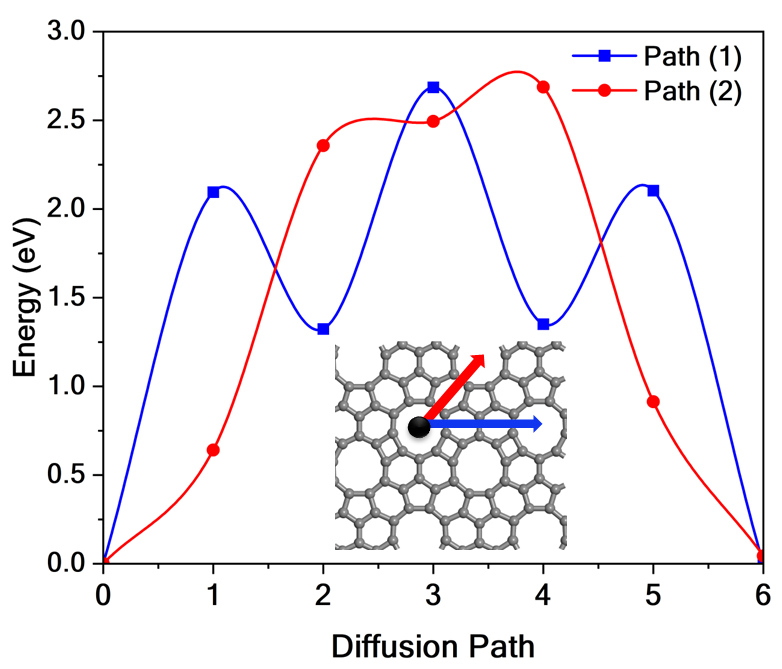}
    \caption{Diffusion energy barrier for Na atom migrates along two paths: the first connecting the enneagonal sites through the naphthylene segments, and the second passing through the distorted hexagonal ring that links adjacent enneagonal pores. The diffusion energy barriers are 2.69 eV for both diffusion paths.}
    \label{fig:neb}
\end{figure}

The surface diffusion of adsorbed Na atoms can lead to clustering, reducing the number of active sites available for hydrogen adsorption. To evaluate the probability of such diffusion, the nudged elastic band (NEB) method \cite{makri2019preconditioning, barzilai1988two, bitzek2006structural} was used, as shown in Fig.~\ref{fig:neb}. This calculation generated seven intermediate images, five of which correspond to transition states.

Two diffusion pathways were considered: connecting neighboring enneagonal pores via the naphthylene segments and traversing the distorted hexagonal ring that links adjacent enneagonal sites. The results show that the energy barrier for Na diffusion is approximately 2.69~eV, indicating a low probability of spontaneous diffusion at room temperature. This value is consistent with those reported for similar systems, such as Ca-decorated biphenylene (2.52~eV) \cite{MA2024129652}, Ti-decorated irida-graphene (5.0~eV) \cite{TAN2024738}, and Sc-decorated biphenylene (3.48~eV) \cite{singh2023highly}, further supporting the structural stability of the Na@TPHE-graphene complex.

This energy barrier is significantly higher than the thermal energy available to a single atom at room temperature (300\,K), which the classical expression can estimate:
\begin{equation}
E_{\text{thermal}} = \frac{3}{2}k_{\text{B}}T
\end{equation}
\noindent where \( k_{\text{B}} = 8.617 \times 10^{-5} \,\text{eV/K} \) is the Boltzmann constant. Substituting the temperature:
\begin{equation}
E_{\text{thermal}} = \frac{3}{2} \times 8.617 \times 10^{-5} \times 300 \approx 0.039 \,\text{eV}.
\end{equation}

Therefore, the diffusion barrier of 2.69\,eV is almost 70 times greater than the thermal energy at 300\,K, indicating that spontaneous diffusion through the pore is thermally inaccessible under ambient conditions and would require elevated temperatures or external activation to occur.

\begin{figure*}
    \centering
    \includegraphics[width=0.8\linewidth]{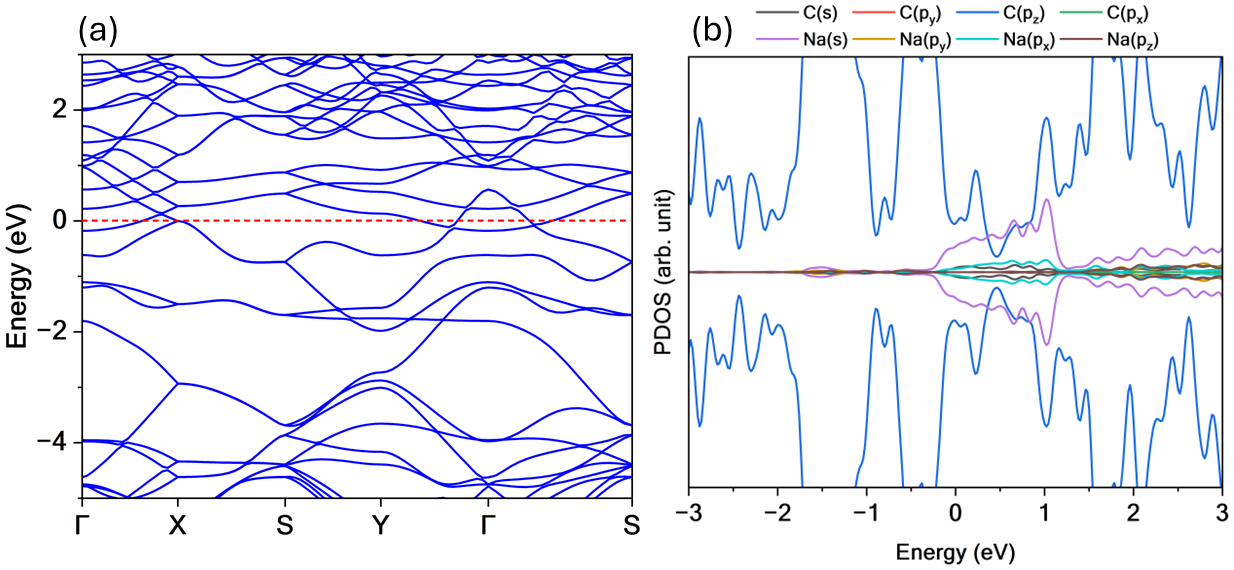}
    \caption{(a) Band structure and (b) PDOS for Na@TPHE-graphene system. The system retains its metallic character with a single band crossing the Fermi level and reduced PDOS in this region.}
    \label{fig:band+dos_Na@TPHE}
\end{figure*}

To investigate the impact of sodium adsorption on the electronic structure of TPHE-graphene, we analyzed the band structure and PDOS for the Na@TPHE-graphene system, as shown in Fig.~\ref{fig:band+dos_Na@TPHE}. The results indicate that the metallic character of the monolayer is preserved, with a single band crossing the Fermi level. The highly dispersive nature of the occupied bands near the Fermi level is also maintained, with a noticeable reduction in the PDOS in this region, suggesting delocalized electronic states.

Several additional bands appear in the conduction band (CB), predominantly associated with Na($s$) and Na($p$) orbitals, as evidenced in the PDOS. Moreover, tilted cone-like crossings are observed near the Fermi level along the \( \Gamma \rightarrow X \), \( Y \rightarrow \Gamma \), and \( \Gamma \rightarrow S \) directions, which may facilitate anisotropic charge transport. Regarding orbital composition, the electronic states are primarily derived from C($p_z$) orbitals, with negligible contributions from other carbon states. Notably, Na($s$) orbitals contribute significantly near the Fermi level, highlighting the role of sodium in modifying the local electronic environment.

\subsection{Hydrogen storage properties of TPHE-graphene}

\begin{figure*}
    \centering
    \includegraphics[width=1\linewidth]{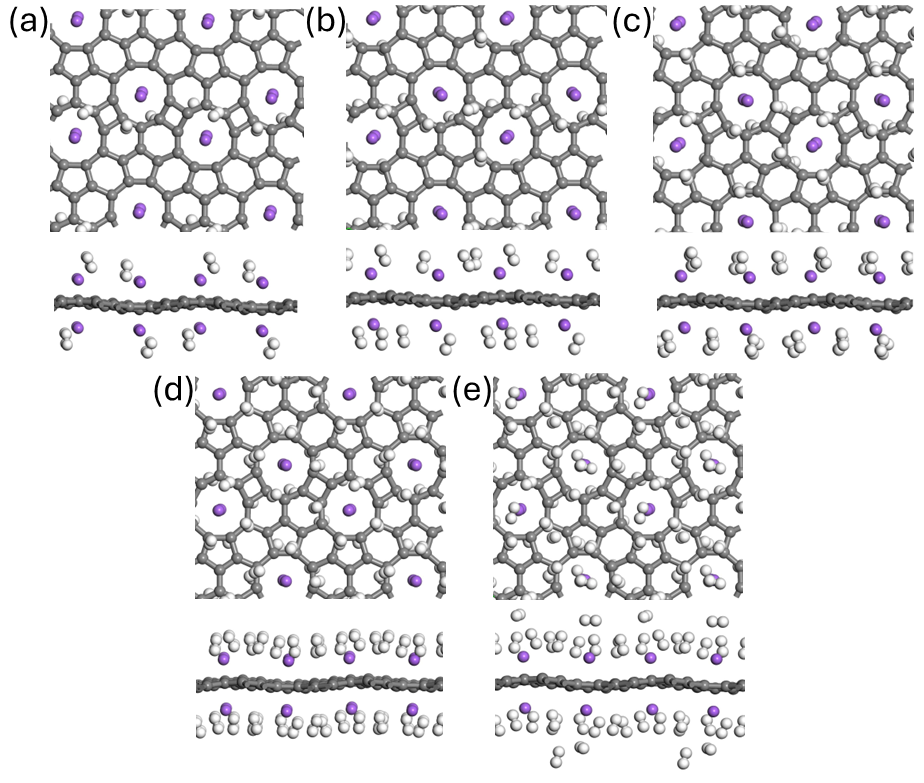}
    \caption{H$_2$ saturation on Na@TPHE-graphene pathway, where (a), (b), (c), (d), and (e) denote Na@TPHE-graphene + 4H$_2$, Na@TPHE-graphene + 8H$_2$, Na@TPHE-graphene + 12H$_2$, Na@TPHE-graphene + 16H$_2$, and Na@TPHE-graphene + 20H$_2$ systems, respectively. Stepwise adsorption of 4 to 20 
    H$_2$ molecules (up to 5 per Na) reveals the gradual saturation behavior of the TPHE-graphene unit cell.}
    \label{fig:H2_saturation}
\end{figure*}

We now evaluate the hydrogen storage potential of the Na@TPHE-graphene system. For this purpose, we sequentially adsorbed one H$_2$ molecule per Na adatom, generating a stepwise series of configurations containing 4, 8, 12, 16, and 20 H$_2$ molecules per TPHE-graphene unit cell, as illustrated in Fig.~\ref{fig:H2_saturation}. The fully saturated configuration accommodates 20 H$_2$ molecules, corresponding to five H$_2$ molecules per Na adatom. In this figure, the H$_2$ saturation on Na@TPHE-graphene pathway, where \ref{fig:H2_saturation}(a), \ref{fig:H2_saturation}(b), \ref{fig:H2_saturation}(c), \ref{fig:H2_saturation}(d), and \ref{fig:H2_saturation}(e) denote Na@TPHE-graphene + 4H$_2$, Na@TPHE-graphene + 8H$_2$, Na@TPHE-graphene + 12H$_2$, Na@TPHE-graphene + 16H$_2$, and Na@TPHE-graphene + 20H$_2$ systems, respectively. Stepwise adsorption of 4 to 20 H$_2$ molecules (up to 5 per Na) reveals the gradual saturation behavior of the TPHE-graphene unit cell.

Key parameters for each adsorption stage—including the average adsorption energy ($E_{\text{ads}}$), consecutive adsorption energy ($E_{\text{con}}$), hydrogen adsorption capacity (HAC), average H--H bond length ($R_{\text{H--H}}$), and estimated desorption temperature ($T_{\text{des}}$), are summarized in Table~\ref{tbl1}.

The average adsorption energy ($E_{\text{ads}}$) remains nearly constant at approximately -0.23~eV up to the 12H$_2$ configuration, indicating favorable and stable interactions during the initial stages of hydrogen adsorption. Beyond this point, a gradual decline in adsorption strength is observed, with $E_{\text{ads}}$ decreasing to -0.21~eV for the 16H$_2$ configuration and reaching -0.18~eV at full saturation (20H$_2$). This trend suggests that the physisorption mechanism becomes progressively weaker as the system approaches its hydrogen storage capacity limit.

Further insight into the energetic cost of incorporating additional H$_2$ molecules is provided by the consecutive adsorption energy ($E_{\text{con}}$). This parameter reaches a peak value of -0.26~eV for the 8H$_2$ configuration, followed by a gradual decline to -0.07~eV at full coverage. Such behavior reflects the progressive saturation of active sites and the increasing repulsive interactions among neighboring adsorbed molecules.

Regarding the hydrogen uptake performance, the hydrogen adsorption capacity (HAC) increases linearly with coverage, achieving a maximum of 9.52~wt\% at full saturation, surpassing the U.S. DOE's target for practical storage materials. Throughout all configurations, the average H--H bond length ($R_{\text{H--H}}$) remains close to that of a free H$_2$ molecule (0.75~\AA), fluctuating only slightly between 0.76 and 0.77~\AA. These values reinforce the physisorption nature of the interaction.

As for thermal behavior, the desorption temperature ($T_{\text{des}}$) remains within a practical range of 243--312~K, enabling hydrogen release under near-ambient conditions. The highest $T_{\text{des}}$ of 312~K is observed for the 8H$_2$ configuration, which coincides with the strongest $E_{\text{con}}$ value.

\begin{figure}
    \centering
    \includegraphics[width=0.6\linewidth]{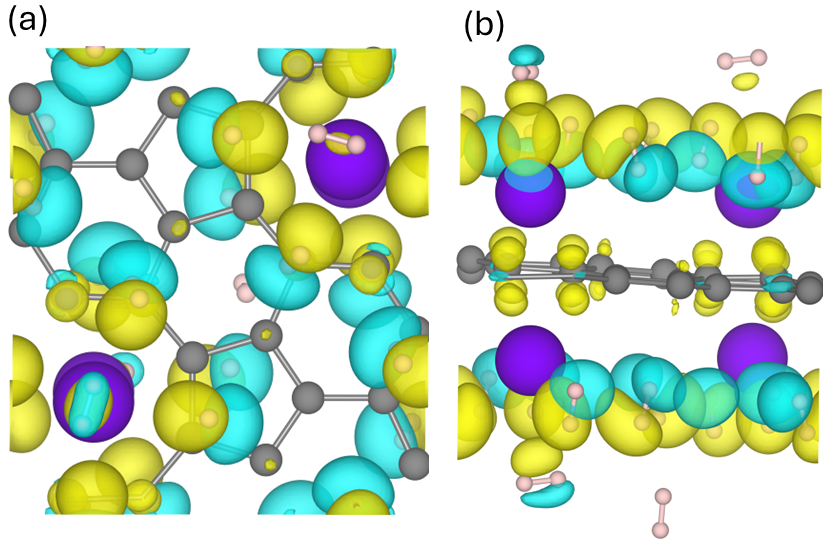}
    \caption{(a) Top and (b) side views of the CDD map for Na@TPHE-graphene + 20H$_2$. The yellow (blue) regions indicate charge accumulation (depletion).}
    \label{fig:CDD_H2}
\end{figure}

To examine the interaction between H$_2$ molecules with the substrate, we analyzed the CDD map, as illustrated in Fig.~\ref{fig:CDD_H2}. The top (Fig. \ref{fig:CDD_H2}(a)) and side (Fig. \ref{fig:CDD_H2}(b)) views of the CDD map reveal that adsorption is primarily driven by polarization effects, with distinct charge accumulation and depletion regions appearing on each hydrogen atom. This redistribution indicates that H$_2$ molecules develop induced dipoles in the presence of the substrate, characteristic of physisorption.

The observed behavior aligns well with the computed adsorption energies ($E_{\text{ads}}$), which fall within the physisorption regime. To further validate this mechanism, a Bader charges were obtained, revealing a net charge transfer of approximately +0.34~$|e|$ for the entire Na@TPHE-graphene system and only -0.02~$|e|$ per H$_2$ molecule. These results confirm that the interaction involves minimal electron transfer, reinforcing the conclusion that induced dipole interactions dominate the H$_2$ adsorption process.

\begin{figure*}
    \centering
    \includegraphics[width=1\linewidth]{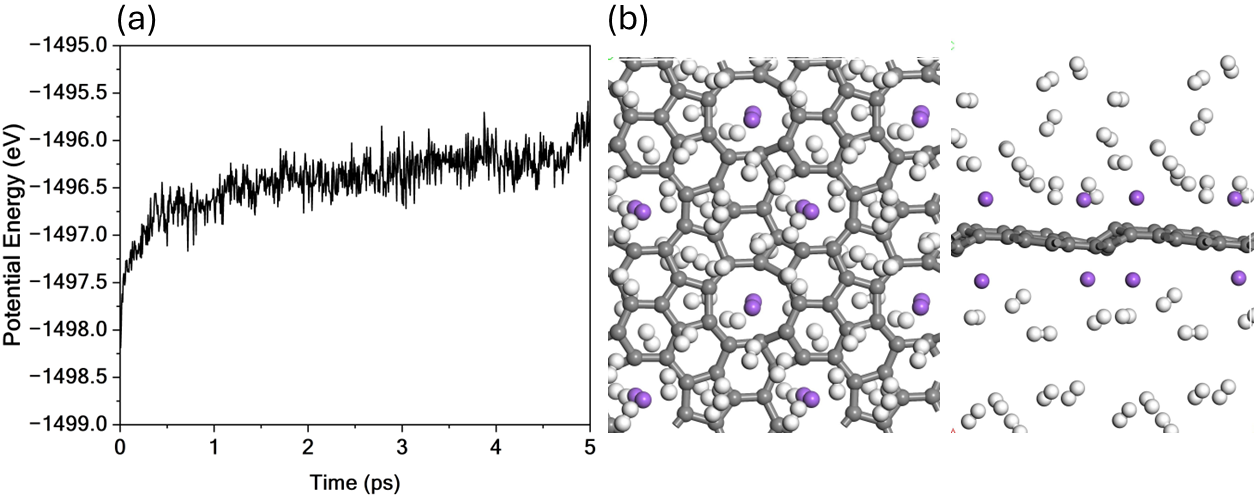}
    \caption{MD simulation results for Na@TPHE-graphene + 20H$_2$ system at 300 K. (a) Time evolution of the potential energy and (b) final system configuration for Na@TPHE-graphene + 20H$_2$ system. High energy fluctuations indicate H$_2$ desorption events, demonstrating the reversible storage capability of Na@TPHE-graphene near ambient conditions.}
    \label{fig:md+h2}
\end{figure*}

To evaluate the thermal stability of the Na@TPHE-graphene system in the presence of hydrogen, MD simulations were conducted for the Na@TPHE-graphene + 20H$_2$ configuration at 300~K over a period of 5~ps, as illustrated in Fig.~\ref{fig:md+h2}. The potential energy profile displays several abrupt fluctuations throughout the simulation, which are attributed to the desorption of H$_2$ molecules. These events provide direct evidence of reversible hydrogen storage, with release occurring under near-ambient conditions.

Despite the dynamic nature of hydrogen desorption, the substrate's structural integrity remains intact throughout the simulation. No atomic rearrangement or bond breaking is observed, and the Na adatoms remain firmly anchored at their preferred adsorption sites, exhibiting only minimal displacements. These findings confirm that the Na-decorated TPHE-graphene maintains its structural robustness while enabling thermally activated hydrogen release.

\begin{table*}[!ht]
\caption{Adsorption energy ($\text{E}_\text{ads}$), hydrogen adsorption capacity (HAC), consecutive adsorption energy (E$_\text{con}$) average H--H bond length (R$_\text{H-H}$), and desorption temperature (T$_\text{des}$) 
for Na@TPHE-graphene + $n$H$_2$ ($n = 2$, 4, 6, 8).}
\label{tbl1}
\centering
\begin{tabular*}{\linewidth}{@{\extracolsep{\fill}} lccccc @{}}
\toprule
\textbf{System} & \textbf{$\text{E}_\text{ads}$ (eV)} & \textbf{E$_\text{con}$ (eV)} & \textbf{HAC (wt\%)} & \textbf{R$_\text{H-H}$ (\AA)} & \textbf{T$_\text{des}$ (K)} \\
\midrule
\textbf{Na@TPHE + 4H$_2$}    & $-0.23$   & $-0.23$ & 2.06  & 0.76 & 292.17 \\
\textbf{Na@TPHE + 8H$_2$}    & $-0.23$   & $-0.26$ & 4.04  & 0.77 & 312.41 \\
\textbf{Na@TPHE + 12H$_2$}    & $-0.23$   & $-0.24$ & 5.94  & 0.77 & 310.65 \\
\textbf{Na@TPHE + 16H$_2$}    & $-0.21$   & $-0.15$ & 7.77  & 0.76 & 280.81 \\
\textbf{Na@TPHE + 20H$_2$}    & $-0.18$   & $-0.07$ & 9.52  & 0.76 & 243.47 \\
\bottomrule
\end{tabular*}
\end{table*}

Na@TPHE-graphene emerges as an up-and-coming candidate among recently proposed hydrogen storage systems due to its optimal combination of adsorption capacity, binding energy, and desorption temperature. In its fully saturated state with 20 adsorbed H$_2$ molecules, the system reaches a hydrogen adsorption capacity (HAC) of 9.52~wt\%, outperforming several benchmark materials such as Na@B$_7$N$_5$ (7.70~wt\%), Na@Irida-graphene (7.82~wt\%), and Na@Graphdiyne (7.7~wt\%).

Compared to other high-performing systems, including K@BP-Biphenylene (8.27~wt\%) and Li@POG-B$_4$C$_2$N$_3$ (8.35~wt\%), Na@TPHE-graphene still exhibits superior storage capacity. Although materials like Ca@DTT-graphene (11.7~wt\%) and Li@$\beta_{12}$-borophene (11.59~wt\%) report even higher capacities, these systems often rely on heavier dopants or lack a thorough evaluation of the desorption properties, an essential factor for practical applications. In contrast, Na@TPHE-graphene balances high capacity with reversible adsorption behavior under near-ambient conditions, reinforcing its potential as a viable hydrogen storage platform.

Regarding adsorption energetics, Na@TPHE-graphene exhibits a binding energy of 0.18~eV per H$_2$, which falls within the optimal range (0.15--0.25~eV) for physisorption-based hydrogen storage systems. This energy is sufficiently high to stabilize hydrogen molecules under moderate pressure yet low enough to allow for desorption without excessive thermal input. 

The value is comparable to those reported for Na@B$_7$N$_5$ (0.20~eV) and Li@POG-B$_4$C$_2$N$_3$ (0.19~eV) and significantly higher than that of Na@IGP-SiC (0.10~eV) and NLi$_4$@Phosphorene (0.11~eV), which may exhibit insufficient binding strength for ambient-condition storage. Although systems like Mg@Me-C$_8$B$_5$ show stronger adsorption (0.29~eV), their practical use is limited by a lower HAC (5.74~wt\%) and elevated desorption temperature (370~K). These comparisons highlight the favorable balance achieved by Na@TPHE-graphene between adsorption strength and reversibility, key attributes for viable hydrogen storage materials.

Another key parameter, the desorption temperature ($T_{\text{des}}$) of 243~K, positions Na@TPHE-graphene among the most balanced candidates for practical hydrogen storage. This temperature is low enough to allow hydrogen release under near-ambient or mildly elevated conditions, yet sufficiently high to prevent premature desorption during storage.

In comparison, materials such as Na@Irida-graphene ($T_{\text{des}} = 195$~K) and NLi$_4$@Phosphorene ($T_{\text{des}} = 82$~K) may suffer from hydrogen loss under mild conditions, limiting their applicability. Conversely, while systems like Mg@Me-C$_8$B$_5$ (370~K) and Ca@DTT-graphene (284~K) offer strong thermal stability, they require substantial heating to initiate hydrogen release, which can reduce overall energy efficiency. Na@TPHE-graphene thus offers a favorable compromise between storage safety and energy-effective release.

\begin{table*}[!ht]
\label{table:adsorption}
\centering
\caption{Number of adsorbed H$_2$ molecules ($n$), absolute adsorption energy per H$_2$ (|$\text{E}_\text{ads}$|), Hydrogen Adsorption Capacity (HAC), and Desorption temperature (T$_\text{des}$) associated with configurations 
exhibiting complete H$_2$ coverage configurations in recently documented systems.} 
\begin{tabular*}{\linewidth}{@{\extracolsep{\fill}} lcccc @{}}
\toprule
\textbf{System}                      & \textbf{n} & \textbf{|$\text{E}_\text{ads}$| (eV)} & \textbf{HAC (wt\%)} & \textbf{T$_\text{des}$ (K)} \\
\midrule
\textbf{Na@TPHE-graphene} (this work) & 20 & 0.18 & 9.52 & 243 \\
\textbf{Na@B$_7$N$_5$} \cite{LIU2025105802} & 32  & 0.20  & 7.70  & 257  \\ 
\textbf{Na@Irida-graphene} \cite{duan2024reversible} & 32 & 0.14 & 7.82 & 195 \\
\textbf{Na@Graphdiyne} \cite{wang2020lithium} & 5 & 0.25 & 7.7 & - \\
\textbf{Na@IGP-SiC} \cite{MARTINS202498} & 48 & 0.10 & 6.78 & 148 \\
\textbf{K@BP-Biphenylene} \cite{djebablia2024metal} & 32 & 0.14 & 8.27 & - \\
\textbf{Mg@Me-C$_8$B$_5$} \cite{gong2024me}  & 3  & 0.29  & 5.74  & 370 \\ 
\textbf{NLi$_4$@Phosphorene} \cite{boubkri2024computational} & 30 & 0.11 & 6.8 & 82 \\
\textbf{Ca@DTT-graphene} \cite{guo2024dtt} & 5 & 0.22 & 11.7 & 284 \\
\textbf{Li@POG-B$_4$C$_2$N$_3$} \cite{chen2025penta} & 10 & 0.19 & 8.35 & 245 \\
\textbf{Li@$\beta_{12}$-borophene} \cite{KUMAR2024510}  & 4 & 0.10 & 11.59 & - \\ 

\bottomrule
\end{tabular*}
\end{table*}

\begin{figure*}
    \centering
    \includegraphics[width=0.8\linewidth]{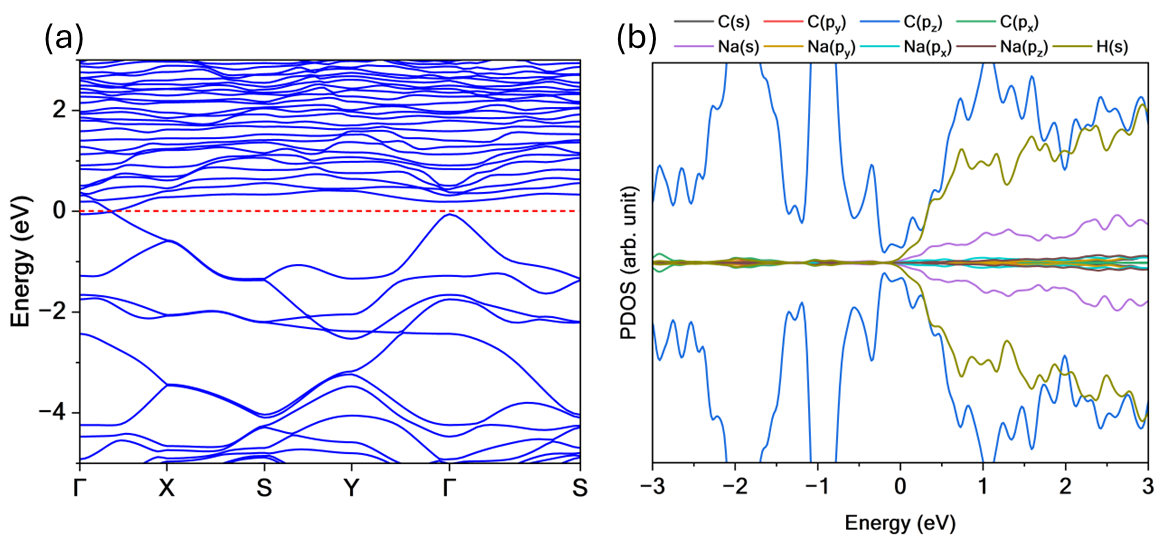}
    \caption{(a) Band structure and (b) PDOS for Na@TPHE-graphene + 20H$_2$. While the valence band retains features of the pristine and Na-decorated systems, the conduction band displays new bands and increased PDOS, 
    mainly due to H\textsubscript{2} and Na(\textit{s}) states. A tilted cone-like crossing at the Fermi level (red dashed line), along with reduced DOS, shows a metallic to semi-metallic transition.}
    \label{fig:band+H2}
\end{figure*}

To assess the electronic impact of H$_2$ adsorption on Na@TPHE-graphene, we examined the electronic band structure and PDOS for the Na@TPHE-graphene + 20H$_2$ system, as shown in Fig.~\ref{fig:band+H2}. In the valence band, the overall electronic features remain largely unchanged compared to both the pristine and Na-decorated TPHE-graphene cases. 

This observation is corroborated by the PDOS analysis, which shows negligible contributions from both Na and H$_2$ in the valence band region. As with the undeformed system, the valence states are predominantly composed of carbon $p_z$ orbitals, highlighting the preservation of the $\pi$-bonding network upon hydrogen adsorption.

In contrast, the conduction band exhibits noticeable modifications upon H$_2$ adsorption. Several new bands emerge, accompanied by a marked increase in the density of states. The PDOS reveals that H$_2$ molecules contribute significantly to the electronic states in this region, followed by strong contributions from Na $s$ orbitals.

Carbon $p_z$ orbitals remain dominant in the conduction band, underscoring their continued relevance to the overall electronic structure. These findings suggest that, while the valence band remains largely unaffected, the conduction band is more sensitive to hydrogen adsorption, potentially influencing the material’s charge transport characteristics. 

A tilted cone-like crossing is observed at the Fermi level, accompanied by a reduction in the density of states at this energy. This feature indicates a transition from metallic to semi-metallic behavior after full H\textsubscript{2} coverage.

\begin{figure}[!ht]
    \centering
    \includegraphics[width=0.8\linewidth]{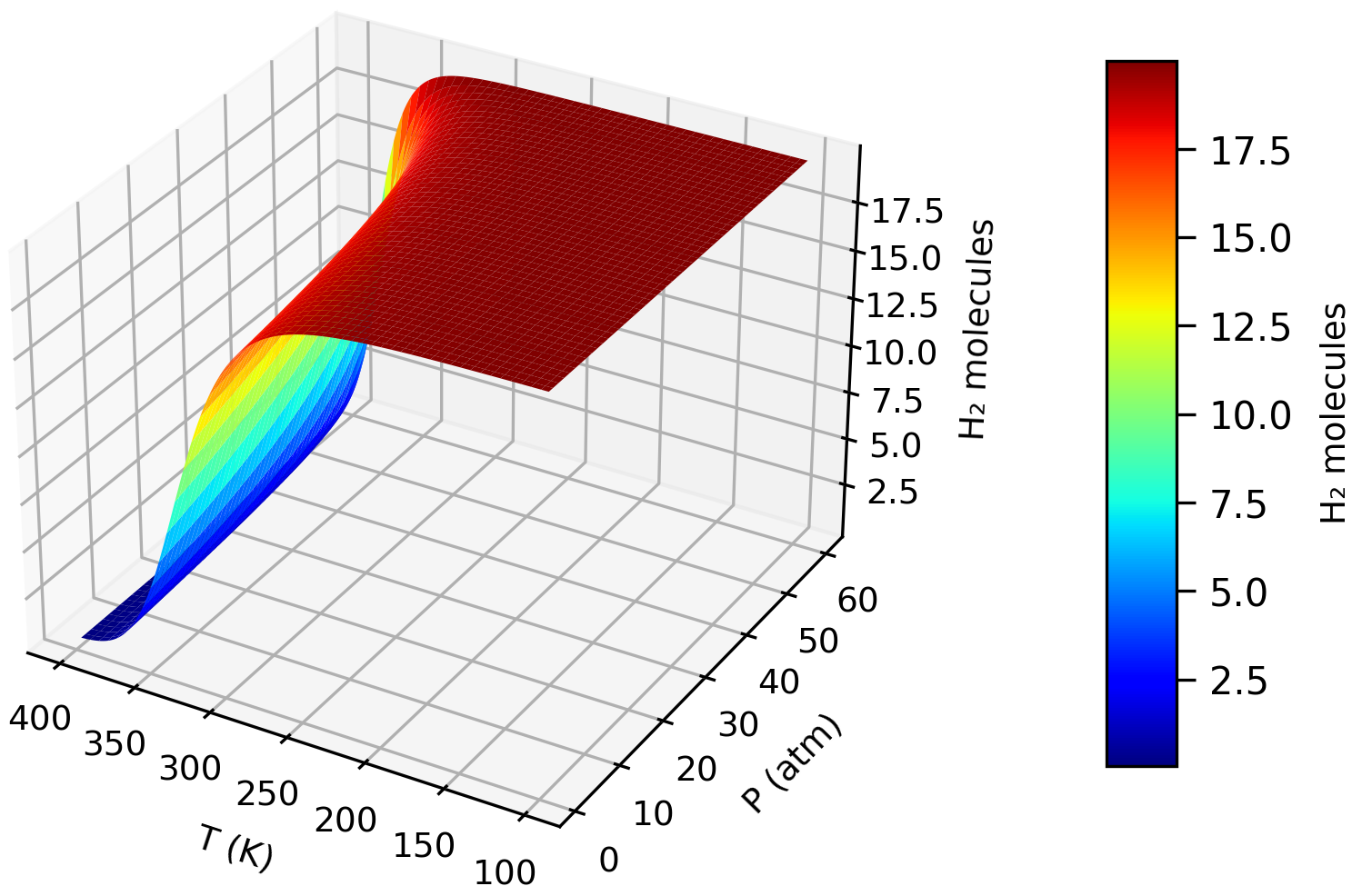}
    \caption{The average number of adsorbed H$_2$ on Na@TPHE-graphene at various temperatures (T) and pressures (P).}
    \label{fig:12}
\end{figure}

When analyzing the behavior of hydrogen molecules, it is essential to consider that H$_2$ can be efficiently stored under low temperature and high-pressure conditions, while desorption typically occurs at elevated and reduced pressures. We refer to the thermodynamic analysis presented in Fig.~\ref{fig:12} to evaluate the practical feasibility of hydrogen adsorption and release.

In typical operating conditions, adsorption takes place at 30~atm and a temperature of 25\,$^\circ$C, while desorption is triggered by reducing the pressure to 3~atm and raising the temperature to 100\,$^\circ$C. Under these conditions, the number of adsorbed H$_2$ molecules reaches 19.53 during adsorption and drops to 0.16 upon desorption. This difference corresponds to an effective hydrogen storage capacity of 9.25~wt\%, demonstrating the material’s practical viability for reversible hydrogen storage applications.

\section{Conclusions}

In summary, we propose TPHE-graphene as a high-performance platform for hydrogen storage via sodium decoration. This novel 2D carbon monolayer, composed of 4-, 5-, 6-, and 9-membered rings, was recently identified through a high-throughput structure search. Its dynamical, energetic, and thermal stability were confirmed through phonon dispersion analysis, cohesive energy calculations, and MD simulations.

Na decoration leads to strong chemisorption, characterized by a binding energy of -2.08~eV and significant charge transfer to the monolayer. The electronic structure remains metallic after decoration, and MD simulations confirm that Na atoms remain stably anchored at room temperature. These features ensure the structural robustness of the decorated system under realistic conditions.

Hydrogen adsorption studies reveal that each Na atom can host up to five H$_2$ molecules, resulting in a gravimetric storage capacity of 9.52~wt\%. Adsorption energies lie within the optimal range for physisorption, and thermodynamic simulations indicate reversible H$_2$ release near ambient conditions. A practical hydrogen storage capacity of 9.25~wt\% is achieved under typical working conditions, positioning Na@TPHE-graphene as a competitive and reliable candidate for solid-state hydrogen storage applications.

\section*{Data access statement}
Data supporting the results can be accessed by contacting the corresponding author.

\section*{Conflicts of interest}
The authors declare no conflict of interest.

\section*{Acknowledgements}
The authors acknowledge financial support from several Brazilian funding agencies. This research was funded by the São Paulo Research Foundation (FAPESP) through Grants No. 2022/03959-6, 2024/05087-1, 2013/08293-7, 2020/01144-0, 2024/19996-3, and 2022/16509-9. Additional support was provided by the National Council for Scientific and Technological Development (CNPq), under Grant No. 307213/2021–8. L.A.R.J. gratefully acknowledges funding from FAP-DF through Grants No. 00193.00001808/2022-71, 00193-00001857/2023-95, and the FAPDF-PRONEM initiative (Grant No. 00193.00001247/2021-20), as well as the PDPG-FAPDF-CAPES Centro-Oeste program (Grant No. 00193-00000867/2024-94). Support from CNPq was also received through Grants No. 350176/2022-1, 167745/2023-9, and 444111/2024-7. Computational resources were provided by the Centro Nacional de Processamento de Alto Desempenho em São Paulo (CENAPAD-SP), CENAPAD-RJ (SDumont), and the Molecular Simulations Laboratory at São Paulo State University (UNESP), Bauru campus.

\section*{Funding}
The Article Processing Charge for the publication of this research was funded by the Coordenacao de Aperfeicoamento de Pessoal de Nivel Superior (CAPES), Brazil (ROR identifier: 00x0ma614).

\bibliography{jshort_2024,achemso-demo}

\providecommand{\latin}[1]{#1}
\makeatletter
\providecommand{\doi}
  {\begingroup\let\do\@makeother\dospecials
  \catcode`\{=1 \catcode`\}=2 \doi@aux}
\providecommand{\doi@aux}[1]{\endgroup\texttt{#1}}
\makeatother
\providecommand*\mcitethebibliography{\thebibliography}
\csname @ifundefined\endcsname{endmcitethebibliography}  {\let\endmcitethebibliography\endthebibliography}{}
\begin{mcitethebibliography}{63}
\providecommand*\natexlab[1]{#1}
\providecommand*\mciteSetBstSublistMode[1]{}
\providecommand*\mciteSetBstMaxWidthForm[2]{}
\providecommand*\mciteBstWouldAddEndPuncttrue
  {\def\EndOfBibitem{\unskip.}}
\providecommand*\mciteBstWouldAddEndPunctfalse
  {\let\EndOfBibitem\relax}
\providecommand*\mciteSetBstMidEndSepPunct[3]{}
\providecommand*\mciteSetBstSublistLabelBeginEnd[3]{}
\providecommand*\EndOfBibitem{}
\mciteSetBstSublistMode{f}
\mciteSetBstMaxWidthForm{subitem}{(\alph{mcitesubitemcount})}
\mciteSetBstSublistLabelBeginEnd
  {\mcitemaxwidthsubitemform\space}
  {\relax}
  {\relax}

\bibitem[Pani \latin{et~al.}(2022)Pani, Shirkole, and Mujumdar]{pani2022importance}
Pani,~A.; Shirkole,~S.~S.; Mujumdar,~A.~S. Importance of renewable energy in the fight against global climate change. 2022\relax
\mciteBstWouldAddEndPuncttrue
\mciteSetBstMidEndSepPunct{\mcitedefaultmidpunct}
{\mcitedefaultendpunct}{\mcitedefaultseppunct}\relax
\EndOfBibitem
\bibitem[Oliveira \latin{et~al.}(2021)Oliveira, Beswick, and Yan]{oliveira2021green}
Oliveira,~A.~M.; Beswick,~R.~R.; Yan,~Y. A green hydrogen economy for a renewable energy society. \emph{Current Opinion in Chemical Engineering} \textbf{2021}, \emph{33}, 100701\relax
\mciteBstWouldAddEndPuncttrue
\mciteSetBstMidEndSepPunct{\mcitedefaultmidpunct}
{\mcitedefaultendpunct}{\mcitedefaultseppunct}\relax
\EndOfBibitem
\bibitem[Laimon and Yusaf(2024)Laimon, and Yusaf]{laimon2024towards}
Laimon,~M.; Yusaf,~T. Towards energy freedom: Exploring sustainable solutions for energy independence and self-sufficiency using integrated renewable energy-driven hydrogen system. \emph{Renewable Energy} \textbf{2024}, \emph{222}, 119948\relax
\mciteBstWouldAddEndPuncttrue
\mciteSetBstMidEndSepPunct{\mcitedefaultmidpunct}
{\mcitedefaultendpunct}{\mcitedefaultseppunct}\relax
\EndOfBibitem
\bibitem[Ma \latin{et~al.}(2024)Ma, Zhao, Wang, Li, and Zhou]{ma2024large}
Ma,~N.; Zhao,~W.; Wang,~W.; Li,~X.; Zhou,~H. Large scale of green hydrogen storage: Opportunities and challenges. \emph{International Journal of Hydrogen Energy} \textbf{2024}, \emph{50}, 379--396\relax
\mciteBstWouldAddEndPuncttrue
\mciteSetBstMidEndSepPunct{\mcitedefaultmidpunct}
{\mcitedefaultendpunct}{\mcitedefaultseppunct}\relax
\EndOfBibitem
\bibitem[Dewangan \latin{et~al.}(2022)Dewangan, Mohan, Kumar, Sharma, and Ahn]{dewangan2022comprehensive}
Dewangan,~S.~K.; Mohan,~M.; Kumar,~V.; Sharma,~A.; Ahn,~B. A comprehensive review of the prospects for future hydrogen storage in materials-application and outstanding issues. \emph{International Journal of Energy Research} \textbf{2022}, \emph{46}, 16150--16177\relax
\mciteBstWouldAddEndPuncttrue
\mciteSetBstMidEndSepPunct{\mcitedefaultmidpunct}
{\mcitedefaultendpunct}{\mcitedefaultseppunct}\relax
\EndOfBibitem
\bibitem[Reilly~Jr and Wiswall~Jr(1968)Reilly~Jr, and Wiswall~Jr]{reilly1968reaction}
Reilly~Jr,~J.~J.; Wiswall~Jr,~R.~H. Reaction of hydrogen with alloys of magnesium and nickel and the formation of Mg2NiH4. \emph{Inorganic chemistry} \textbf{1968}, \emph{7}, 2254--2256\relax
\mciteBstWouldAddEndPuncttrue
\mciteSetBstMidEndSepPunct{\mcitedefaultmidpunct}
{\mcitedefaultendpunct}{\mcitedefaultseppunct}\relax
\EndOfBibitem
\bibitem[Rusman and Dahari(2016)Rusman, and Dahari]{rusman2016review}
Rusman,~N.; Dahari,~M. A review on the current progress of metal hydrides material for solid-state hydrogen storage applications. \emph{International Journal of Hydrogen Energy} \textbf{2016}, \emph{41}, 12108--12126\relax
\mciteBstWouldAddEndPuncttrue
\mciteSetBstMidEndSepPunct{\mcitedefaultmidpunct}
{\mcitedefaultendpunct}{\mcitedefaultseppunct}\relax
\EndOfBibitem
\bibitem[Pinjari \latin{et~al.}(2023)Pinjari, Bera, Kapur, and Kjeang]{pinjari2023mechanism}
Pinjari,~S.; Bera,~T.; Kapur,~G.; Kjeang,~E. The mechanism and sorption kinetic analysis of hydrogen storage at room temperature using acid functionalized carbon nanotubes. \emph{International Journal of Hydrogen Energy} \textbf{2023}, \emph{48}, 1930--1942\relax
\mciteBstWouldAddEndPuncttrue
\mciteSetBstMidEndSepPunct{\mcitedefaultmidpunct}
{\mcitedefaultendpunct}{\mcitedefaultseppunct}\relax
\EndOfBibitem
\bibitem[Cardoso \latin{et~al.}(2021)Cardoso, Piquini, Khossossi, and Ahuja]{cardoso2021lithium}
Cardoso,~G.~L.; Piquini,~P.~C.; Khossossi,~N.; Ahuja,~R. Lithium-functionalized boron phosphide nanotubes (BPNTs) as an efficient hydrogen storage carrier. \emph{International Journal of Hydrogen Energy} \textbf{2021}, \emph{46}, 20586--20593\relax
\mciteBstWouldAddEndPuncttrue
\mciteSetBstMidEndSepPunct{\mcitedefaultmidpunct}
{\mcitedefaultendpunct}{\mcitedefaultseppunct}\relax
\EndOfBibitem
\bibitem[Shet \latin{et~al.}(2021)Shet, Priya, Sudhakar, and Tahir]{shet2021review}
Shet,~S.~P.; Priya,~S.~S.; Sudhakar,~K.; Tahir,~M. A review on current trends in potential use of metal-organic framework for hydrogen storage. \emph{International Journal of Hydrogen Energy} \textbf{2021}, \emph{46}, 11782--11803\relax
\mciteBstWouldAddEndPuncttrue
\mciteSetBstMidEndSepPunct{\mcitedefaultmidpunct}
{\mcitedefaultendpunct}{\mcitedefaultseppunct}\relax
\EndOfBibitem
\bibitem[Kim \latin{et~al.}(2024)Kim, So, Muhammad, and Oh]{kim2024comparing}
Kim,~H.; So,~S.~H.; Muhammad,~R.; Oh,~H. Comparing the practical hydrogen storage capacity of porous adsorbents: Activated carbon and metal-organic framework. \emph{International Journal of Hydrogen Energy} \textbf{2024}, \emph{50}, 1616--1625\relax
\mciteBstWouldAddEndPuncttrue
\mciteSetBstMidEndSepPunct{\mcitedefaultmidpunct}
{\mcitedefaultendpunct}{\mcitedefaultseppunct}\relax
\EndOfBibitem
\bibitem[Bishnoi \latin{et~al.}(2024)Bishnoi, Pati, and Sharma]{bishnoi2024architectural}
Bishnoi,~A.; Pati,~S.; Sharma,~P. Architectural design of metal hydrides to improve the hydrogen storage characteristics. \emph{Journal of Power Sources} \textbf{2024}, \emph{608}, 234609\relax
\mciteBstWouldAddEndPuncttrue
\mciteSetBstMidEndSepPunct{\mcitedefaultmidpunct}
{\mcitedefaultendpunct}{\mcitedefaultseppunct}\relax
\EndOfBibitem
\bibitem[Schneemann \latin{et~al.}(2018)Schneemann, White, Kang, Jeong, Wan, Cho, Heo, Prendergast, Urban, Wood, \latin{et~al.} others]{schneemann2018nanostructured}
Schneemann,~A.; White,~J.~L.; Kang,~S.; Jeong,~S.; Wan,~L.~F.; Cho,~E.~S.; Heo,~T.~W.; Prendergast,~D.; Urban,~J.~J.; Wood,~B.~C.; others Nanostructured metal hydrides for hydrogen storage. \emph{Chemical reviews} \textbf{2018}, \emph{118}, 10775--10839\relax
\mciteBstWouldAddEndPuncttrue
\mciteSetBstMidEndSepPunct{\mcitedefaultmidpunct}
{\mcitedefaultendpunct}{\mcitedefaultseppunct}\relax
\EndOfBibitem
\bibitem[Abifarin \latin{et~al.}(2024)Abifarin, Torres, and Lu]{abifarin20242d}
Abifarin,~J.~K.; Torres,~J.~F.; Lu,~Y. 2D materials for enabling hydrogen as an energy vector. \emph{Nano Energy} \textbf{2024}, 109997\relax
\mciteBstWouldAddEndPuncttrue
\mciteSetBstMidEndSepPunct{\mcitedefaultmidpunct}
{\mcitedefaultendpunct}{\mcitedefaultseppunct}\relax
\EndOfBibitem
\bibitem[Qin \latin{et~al.}(2023)Qin, Yang, Xing, Zhang, Zhang, and Wu]{qin2023two}
Qin,~J.; Yang,~Z.; Xing,~F.; Zhang,~L.; Zhang,~H.; Wu,~Z.-S. Two-dimensional mesoporous materials for energy storage and conversion: current status, chemical synthesis and challenging perspectives. \emph{Electrochemical Energy Reviews} \textbf{2023}, \emph{6}, 9\relax
\mciteBstWouldAddEndPuncttrue
\mciteSetBstMidEndSepPunct{\mcitedefaultmidpunct}
{\mcitedefaultendpunct}{\mcitedefaultseppunct}\relax
\EndOfBibitem
\bibitem[Lokesh and Srivastava(2022)Lokesh, and Srivastava]{lokesh2022advanced}
Lokesh,~S.; Srivastava,~R. Advanced two-dimensional materials for green hydrogen generation: strategies toward corrosion resistance seawater electrolysis- review and future perspectives. \emph{Energy \& Fuels} \textbf{2022}, \emph{36}, 13417--13450\relax
\mciteBstWouldAddEndPuncttrue
\mciteSetBstMidEndSepPunct{\mcitedefaultmidpunct}
{\mcitedefaultendpunct}{\mcitedefaultseppunct}\relax
\EndOfBibitem
\bibitem[Zhang \latin{et~al.}(2012)Zhang, Zhao, Bu, He, Zhang, Zhao, and Luo]{zhang2012ultra}
Zhang,~H.; Zhao,~M.; Bu,~H.; He,~X.; Zhang,~M.; Zhao,~L.; Luo,~Y. Ultra-high hydrogen storage capacity of Li-decorated graphyne: A first-principles prediction. \emph{Journal of Applied Physics} \textbf{2012}, \emph{112}\relax
\mciteBstWouldAddEndPuncttrue
\mciteSetBstMidEndSepPunct{\mcitedefaultmidpunct}
{\mcitedefaultendpunct}{\mcitedefaultseppunct}\relax
\EndOfBibitem
\bibitem[Gopalsamy and Subramanian(2014)Gopalsamy, and Subramanian]{gopalsamy2014hydrogen}
Gopalsamy,~K.; Subramanian,~V. Hydrogen storage capacity of alkali and alkaline earth metal ions doped carbon based materials: A DFT study. \emph{international journal of hydrogen energy} \textbf{2014}, \emph{39}, 2549--2559\relax
\mciteBstWouldAddEndPuncttrue
\mciteSetBstMidEndSepPunct{\mcitedefaultmidpunct}
{\mcitedefaultendpunct}{\mcitedefaultseppunct}\relax
\EndOfBibitem
\bibitem[Mohajeri and Shahsavar(2018)Mohajeri, and Shahsavar]{mohajeri2018light}
Mohajeri,~A.; Shahsavar,~A. Light metal decoration on nitrogen/sulfur codoped graphyne: an efficient strategy for designing hydrogen storage media. \emph{Physica E: Low-dimensional Systems and Nanostructures} \textbf{2018}, \emph{101}, 167--173\relax
\mciteBstWouldAddEndPuncttrue
\mciteSetBstMidEndSepPunct{\mcitedefaultmidpunct}
{\mcitedefaultendpunct}{\mcitedefaultseppunct}\relax
\EndOfBibitem
\bibitem[Yadav \latin{et~al.}(2016)Yadav, Dashora, Patel, Miotello, Press, and Kothari]{yadav2016study}
Yadav,~A.; Dashora,~A.; Patel,~N.; Miotello,~A.; Press,~M.; Kothari,~D. Study of 2D MXene Cr2C material for hydrogen storage using density functional theory. \emph{Applied Surface Science} \textbf{2016}, \emph{389}, 88--95\relax
\mciteBstWouldAddEndPuncttrue
\mciteSetBstMidEndSepPunct{\mcitedefaultmidpunct}
{\mcitedefaultendpunct}{\mcitedefaultseppunct}\relax
\EndOfBibitem
\bibitem[Yadav \latin{et~al.}(2015)Yadav, Tam, and Singh]{yadav2015first}
Yadav,~S.; Tam,~J.; Singh,~C.~V. A first principles study of hydrogen storage on lithium decorated two dimensional carbon allotropes. \emph{international journal of hydrogen energy} \textbf{2015}, \emph{40}, 6128--6136\relax
\mciteBstWouldAddEndPuncttrue
\mciteSetBstMidEndSepPunct{\mcitedefaultmidpunct}
{\mcitedefaultendpunct}{\mcitedefaultseppunct}\relax
\EndOfBibitem
\bibitem[Li \latin{et~al.}(2011)Li, Li, Wu, Li, Xia, and Wang]{li2011high}
Li,~C.; Li,~J.; Wu,~F.; Li,~S.-S.; Xia,~J.-B.; Wang,~L.-W. High capacity hydrogen storage in Ca decorated graphyne: a first-principles study. \emph{The Journal of Physical Chemistry C} \textbf{2011}, \emph{115}, 23221--23225\relax
\mciteBstWouldAddEndPuncttrue
\mciteSetBstMidEndSepPunct{\mcitedefaultmidpunct}
{\mcitedefaultendpunct}{\mcitedefaultseppunct}\relax
\EndOfBibitem
\bibitem[Srinivasu and Ghosh(2012)Srinivasu, and Ghosh]{srinivasu2012graphyne}
Srinivasu,~K.; Ghosh,~S.~K. Graphyne and graphdiyne: promising materials for nanoelectronics and energy storage applications. \emph{The Journal of Physical Chemistry C} \textbf{2012}, \emph{116}, 5951--5956\relax
\mciteBstWouldAddEndPuncttrue
\mciteSetBstMidEndSepPunct{\mcitedefaultmidpunct}
{\mcitedefaultendpunct}{\mcitedefaultseppunct}\relax
\EndOfBibitem
\bibitem[Kaewmaraya \latin{et~al.}(2023)Kaewmaraya, Thatsami, Tangpakonsab, Kinkla, Kotmool, Menendez, Aguey-Zinsou, and Hussain]{kaewmaraya2023ultrahigh}
Kaewmaraya,~T.; Thatsami,~N.; Tangpakonsab,~P.; Kinkla,~R.; Kotmool,~K.; Menendez,~C.; Aguey-Zinsou,~K.; Hussain,~T. Ultrahigh hydrogen storage using metal-decorated defected biphenylene. \emph{Applied Surface Science} \textbf{2023}, \emph{629}, 157391\relax
\mciteBstWouldAddEndPuncttrue
\mciteSetBstMidEndSepPunct{\mcitedefaultmidpunct}
{\mcitedefaultendpunct}{\mcitedefaultseppunct}\relax
\EndOfBibitem
\bibitem[Ma \latin{et~al.}(2024)Ma, Sun, Jia, and Wu]{ma2024li}
Ma,~L.-J.; Sun,~Y.; Jia,~J.; Wu,~H.-S. Li-decorated B-doped biphenylene network for reversible hydrogen storage. \emph{Fuel} \textbf{2024}, \emph{357}, 129652\relax
\mciteBstWouldAddEndPuncttrue
\mciteSetBstMidEndSepPunct{\mcitedefaultmidpunct}
{\mcitedefaultendpunct}{\mcitedefaultseppunct}\relax
\EndOfBibitem
\bibitem[Zhang \latin{et~al.}(2024)Zhang, Liu, Guo, Cao, and Wang]{zhang2024decorated}
Zhang,~Y.; Liu,~Z.; Guo,~J.; Cao,~Z.; Wang,~J. Ca-decorated 2D Irida-graphene as a promising hydrogen storage material: A combination of DFT and AIMD study. \emph{International Journal of Hydrogen Energy} \textbf{2024}, \emph{91}, 118--126\relax
\mciteBstWouldAddEndPuncttrue
\mciteSetBstMidEndSepPunct{\mcitedefaultmidpunct}
{\mcitedefaultendpunct}{\mcitedefaultseppunct}\relax
\EndOfBibitem
\bibitem[Zhang and Guo(2024)Zhang, and Guo]{zhang2024li}
Zhang,~Y.-F.; Guo,~J. Li-decorated 2d irida-graphene as a potential hydrogen storage material: A dispersion-corrected density functional theory calculations. \emph{International Journal of Hydrogen Energy} \textbf{2024}, \emph{50}, 1004--1014\relax
\mciteBstWouldAddEndPuncttrue
\mciteSetBstMidEndSepPunct{\mcitedefaultmidpunct}
{\mcitedefaultendpunct}{\mcitedefaultseppunct}\relax
\EndOfBibitem
\bibitem[Mahamiya \latin{et~al.}(2023)Mahamiya, Shukla, and Chakraborty]{mahamiya2023potential}
Mahamiya,~V.; Shukla,~A.; Chakraborty,~B. Potential reversible hydrogen storage in Li-decorated carbon allotrope PAI-Graphene: a first-principles study. \emph{International Journal of Hydrogen Energy} \textbf{2023}, \emph{48}, 37898--37907\relax
\mciteBstWouldAddEndPuncttrue
\mciteSetBstMidEndSepPunct{\mcitedefaultmidpunct}
{\mcitedefaultendpunct}{\mcitedefaultseppunct}\relax
\EndOfBibitem
\bibitem[Laranjeira \latin{et~al.}(2025)Laranjeira, Martins, Ye, Sambrano, and Chen]{LARANJEIRA2025139}
Laranjeira,~J.~A.; Martins,~N.~F.; Ye,~L.; Sambrano,~J.~R.; Chen,~X. Hydrogen storage engineering in PHE-graphene monolayer via potassium (K) decoration. \emph{International Journal of Hydrogen Energy} \textbf{2025}, \emph{123}, 139--149\relax
\mciteBstWouldAddEndPuncttrue
\mciteSetBstMidEndSepPunct{\mcitedefaultmidpunct}
{\mcitedefaultendpunct}{\mcitedefaultseppunct}\relax
\EndOfBibitem
\bibitem[Adithya \latin{et~al.}(2024)Adithya, Jethawa, Ali, and Chakraborty]{adithya2024decorated}
Adithya,~S.; Jethawa,~U.; Ali,~S.~M.; Chakraborty,~B. Ca-decorated holey graphyne for reversible hydrogen storage: Insights from DFT analysis. \emph{International Journal of Hydrogen Energy} \textbf{2024}, \emph{90}, 470--480\relax
\mciteBstWouldAddEndPuncttrue
\mciteSetBstMidEndSepPunct{\mcitedefaultmidpunct}
{\mcitedefaultendpunct}{\mcitedefaultseppunct}\relax
\EndOfBibitem
\bibitem[Chen \latin{et~al.}(2025)Chen, Wang, Martins, Sambrano, and Laranjeira]{chen2025penta}
Chen,~X.; Wang,~J.; Martins,~N.~F.; Sambrano,~J.~R.; Laranjeira,~J.~A. Penta-Octa B4C2N3: A New 2D Material for High-Performance Energy Applications. \emph{Langmuir} \textbf{2025}, \relax
\mciteBstWouldAddEndPunctfalse
\mciteSetBstMidEndSepPunct{\mcitedefaultmidpunct}
{}{\mcitedefaultseppunct}\relax
\EndOfBibitem
\bibitem[Shi \latin{et~al.}(2021)Shi, Li, Li, Ouyang, Zhang, Tang, He, and Zhong]{doi:10.1021/acs.jpclett.1c03193}
Shi,~X.; Li,~S.; Li,~J.; Ouyang,~T.; Zhang,~C.; Tang,~C.; He,~C.; Zhong,~J. High-Throughput Screening of Two-Dimensional Planar sp2 Carbon Space Associated with a Labeled Quotient Graph. \emph{The Journal of Physical Chemistry Letters} \textbf{2021}, \emph{12}, 11511--11519, PMID: 34797680\relax
\mciteBstWouldAddEndPuncttrue
\mciteSetBstMidEndSepPunct{\mcitedefaultmidpunct}
{\mcitedefaultendpunct}{\mcitedefaultseppunct}\relax
\EndOfBibitem
\bibitem[Perdew \latin{et~al.}(1996)Perdew, Burke, and Ernzerhof]{PhysRevLett.77.3865}
Perdew,~J.~P.; Burke,~K.; Ernzerhof,~M. Generalized Gradient Approximation Made Simple. \emph{Phys. Rev. Lett.} \textbf{1996}, \emph{77}, 3865--3868\relax
\mciteBstWouldAddEndPuncttrue
\mciteSetBstMidEndSepPunct{\mcitedefaultmidpunct}
{\mcitedefaultendpunct}{\mcitedefaultseppunct}\relax
\EndOfBibitem
\bibitem[Ernzerhof and Scuseria(1999)Ernzerhof, and Scuseria]{ernzerhof1999assessment}
Ernzerhof,~M.; Scuseria,~G.~E. Assessment of the Perdew--Burke--Ernzerhof exchange-correlation functional. \emph{The Journal of chemical physics} \textbf{1999}, \emph{110}, 5029--5036\relax
\mciteBstWouldAddEndPuncttrue
\mciteSetBstMidEndSepPunct{\mcitedefaultmidpunct}
{\mcitedefaultendpunct}{\mcitedefaultseppunct}\relax
\EndOfBibitem
\bibitem[Bl\"ochl(1994)]{PhysRevB.50.17953}
Bl\"ochl,~P.~E. Projector augmented-wave method. \emph{Phys. Rev. B} \textbf{1994}, \emph{50}, 17953--17979\relax
\mciteBstWouldAddEndPuncttrue
\mciteSetBstMidEndSepPunct{\mcitedefaultmidpunct}
{\mcitedefaultendpunct}{\mcitedefaultseppunct}\relax
\EndOfBibitem
\bibitem[Kresse and Hafner(1993)Kresse, and Hafner]{kresse1993ab}
Kresse,~G.; Hafner,~J. Ab initio molecular dynamics for liquid metals. \emph{Physical review B} \textbf{1993}, \emph{47}, 558\relax
\mciteBstWouldAddEndPuncttrue
\mciteSetBstMidEndSepPunct{\mcitedefaultmidpunct}
{\mcitedefaultendpunct}{\mcitedefaultseppunct}\relax
\EndOfBibitem
\bibitem[Kresse and Furthm{\"u}ller(1996)Kresse, and Furthm{\"u}ller]{kresse1996efficient}
Kresse,~G.; Furthm{\"u}ller,~J. Efficient iterative schemes for ab initio total-energy calculations using a plane-wave basis set. \emph{Physical review B} \textbf{1996}, \emph{54}, 11169\relax
\mciteBstWouldAddEndPuncttrue
\mciteSetBstMidEndSepPunct{\mcitedefaultmidpunct}
{\mcitedefaultendpunct}{\mcitedefaultseppunct}\relax
\EndOfBibitem
\bibitem[Grimme(2006)]{grimme2006semiempirical}
Grimme,~S. Semiempirical GGA-type density functional constructed with a long-range dispersion correction. \emph{Journal of computational chemistry} \textbf{2006}, \emph{27}, 1787--1799\relax
\mciteBstWouldAddEndPuncttrue
\mciteSetBstMidEndSepPunct{\mcitedefaultmidpunct}
{\mcitedefaultendpunct}{\mcitedefaultseppunct}\relax
\EndOfBibitem
\bibitem[Hourahine \latin{et~al.}(2020)Hourahine, Aradi, Blum, Bonafe, Buccheri, Camacho, Cevallos, Deshaye, Dumitric{\u{a}}, Dominguez, \latin{et~al.} others]{hourahine2020dftb+}
Hourahine,~B.; Aradi,~B.; Blum,~V.; Bonafe,~F.; Buccheri,~A.; Camacho,~C.; Cevallos,~C.; Deshaye,~M.; Dumitric{\u{a}},~T.; Dominguez,~A.; others DFTB+, a software package for efficient approximate density functional theory based atomistic simulations. \emph{The Journal of chemical physics} \textbf{2020}, \emph{152}\relax
\mciteBstWouldAddEndPuncttrue
\mciteSetBstMidEndSepPunct{\mcitedefaultmidpunct}
{\mcitedefaultendpunct}{\mcitedefaultseppunct}\relax
\EndOfBibitem
\bibitem[Gaus \latin{et~al.}(2013)Gaus, Goez, and Elstner]{doi:10.1021/ct300849w}
Gaus,~M.; Goez,~A.; Elstner,~M. Parametrization and Benchmark of DFTB3 for Organic Molecules. \emph{Journal of Chemical Theory and Computation} \textbf{2013}, \emph{9}, 338--354, PMID: 26589037\relax
\mciteBstWouldAddEndPuncttrue
\mciteSetBstMidEndSepPunct{\mcitedefaultmidpunct}
{\mcitedefaultendpunct}{\mcitedefaultseppunct}\relax
\EndOfBibitem
\bibitem[Caldeweyher \latin{et~al.}(2017)Caldeweyher, Bannwarth, and Grimme]{caldeweyher2017extension}
Caldeweyher,~E.; Bannwarth,~C.; Grimme,~S. Extension of the D3 dispersion coefficient model. \emph{The Journal of chemical physics} \textbf{2017}, \emph{147}\relax
\mciteBstWouldAddEndPuncttrue
\mciteSetBstMidEndSepPunct{\mcitedefaultmidpunct}
{\mcitedefaultendpunct}{\mcitedefaultseppunct}\relax
\EndOfBibitem
\bibitem[Berendsen \latin{et~al.}(1984)Berendsen, Postma, Van~Gunsteren, DiNola, and Haak]{berendsen1984molecular}
Berendsen,~H.~J.; Postma,~J.~v.; Van~Gunsteren,~W.~F.; DiNola,~A.; Haak,~J.~R. Molecular dynamics with coupling to an external bath. \emph{The Journal of chemical physics} \textbf{1984}, \emph{81}, 3684--3690\relax
\mciteBstWouldAddEndPuncttrue
\mciteSetBstMidEndSepPunct{\mcitedefaultmidpunct}
{\mcitedefaultendpunct}{\mcitedefaultseppunct}\relax
\EndOfBibitem
\bibitem[Durbin and Malardier-Jugroot(2013)Durbin, and Malardier-Jugroot]{durbin2013review}
Durbin,~D.~J.; Malardier-Jugroot,~C. Review of hydrogen storage techniques for on board vehicle applications. \emph{International journal of hydrogen energy} \textbf{2013}, \emph{38}, 14595--14617\relax
\mciteBstWouldAddEndPuncttrue
\mciteSetBstMidEndSepPunct{\mcitedefaultmidpunct}
{\mcitedefaultendpunct}{\mcitedefaultseppunct}\relax
\EndOfBibitem
\bibitem[Alhameedi \latin{et~al.}(2019)Alhameedi, Karton, Jayatilaka, and Hussain]{alhameedi2019metal}
Alhameedi,~K.; Karton,~A.; Jayatilaka,~D.; Hussain,~T. Metal functionalized inorganic nano-sheets as promising materials for clean energy storage. \emph{Applied Surface Science} \textbf{2019}, \emph{471}, 887--892\relax
\mciteBstWouldAddEndPuncttrue
\mciteSetBstMidEndSepPunct{\mcitedefaultmidpunct}
{\mcitedefaultendpunct}{\mcitedefaultseppunct}\relax
\EndOfBibitem
\bibitem[Hashmi \latin{et~al.}(2017)Hashmi, Farooq, Khan, Son, and Hong]{hashmi2017ultra}
Hashmi,~A.; Farooq,~M.~U.; Khan,~I.; Son,~J.; Hong,~J. Ultra-high capacity hydrogen storage in a Li decorated two-dimensional C 2 N layer. \emph{Journal of Materials Chemistry A} \textbf{2017}, \emph{5}, 2821--2828\relax
\mciteBstWouldAddEndPuncttrue
\mciteSetBstMidEndSepPunct{\mcitedefaultmidpunct}
{\mcitedefaultendpunct}{\mcitedefaultseppunct}\relax
\EndOfBibitem
\bibitem[Mouhat and Coudert(2014)Mouhat, and Coudert]{PhysRevB.90.224104}
Mouhat,~F.; Coudert,~F. m. c.-X. Necessary and sufficient elastic stability conditions in various crystal systems. \emph{Phys. Rev. B} \textbf{2014}, \emph{90}, 224104\relax
\mciteBstWouldAddEndPuncttrue
\mciteSetBstMidEndSepPunct{\mcitedefaultmidpunct}
{\mcitedefaultendpunct}{\mcitedefaultseppunct}\relax
\EndOfBibitem
\bibitem[Ying \latin{et~al.}(2020)Ying, Fan, Zhu, Luo, and Huang]{doi:10.1021/acs.jpcc.9b09593}
Ying,~Y.; Fan,~K.; Zhu,~S.; Luo,~X.; Huang,~H. Theoretical Investigation of Monolayer RhTeCl Semiconductors as Photocatalysts for Water Splitting. \emph{The Journal of Physical Chemistry C} \textbf{2020}, \emph{124}, 639--646\relax
\mciteBstWouldAddEndPuncttrue
\mciteSetBstMidEndSepPunct{\mcitedefaultmidpunct}
{\mcitedefaultendpunct}{\mcitedefaultseppunct}\relax
\EndOfBibitem
\bibitem[Makri \latin{et~al.}(2019)Makri, Ortner, and Kermode]{makri2019preconditioning}
Makri,~S.; Ortner,~C.; Kermode,~J.~R. A preconditioning scheme for minimum energy path finding methods. \emph{The Journal of Chemical Physics} \textbf{2019}, \emph{150}, 094109\relax
\mciteBstWouldAddEndPuncttrue
\mciteSetBstMidEndSepPunct{\mcitedefaultmidpunct}
{\mcitedefaultendpunct}{\mcitedefaultseppunct}\relax
\EndOfBibitem
\bibitem[Barzilai and Borwein(1988)Barzilai, and Borwein]{barzilai1988two}
Barzilai,~J.; Borwein,~J.~M. Two-point step size gradient methods. \emph{IMA journal of numerical analysis} \textbf{1988}, \emph{8}, 141--148\relax
\mciteBstWouldAddEndPuncttrue
\mciteSetBstMidEndSepPunct{\mcitedefaultmidpunct}
{\mcitedefaultendpunct}{\mcitedefaultseppunct}\relax
\EndOfBibitem
\bibitem[Bitzek \latin{et~al.}(2006)Bitzek, Koskinen, G{\"a}hler, Moseler, and Gumbsch]{bitzek2006structural}
Bitzek,~E.; Koskinen,~P.; G{\"a}hler,~F.; Moseler,~M.; Gumbsch,~P. Structural relaxation made simple. \emph{Physical review letters} \textbf{2006}, \emph{97}, 170201\relax
\mciteBstWouldAddEndPuncttrue
\mciteSetBstMidEndSepPunct{\mcitedefaultmidpunct}
{\mcitedefaultendpunct}{\mcitedefaultseppunct}\relax
\EndOfBibitem
\bibitem[Ma \latin{et~al.}(2024)Ma, Sun, Jia, and Wu]{MA2024129652}
Ma,~L.-J.; Sun,~Y.; Jia,~J.; Wu,~H.-S. Li-decorated B-doped biphenylene network for reversible hydrogen storage. \emph{Fuel} \textbf{2024}, \emph{357}, 129652\relax
\mciteBstWouldAddEndPuncttrue
\mciteSetBstMidEndSepPunct{\mcitedefaultmidpunct}
{\mcitedefaultendpunct}{\mcitedefaultseppunct}\relax
\EndOfBibitem
\bibitem[Tan \latin{et~al.}(2024)Tan, Tao, Ouyang, and Peng]{TAN2024738}
Tan,~Y.; Tao,~X.; Ouyang,~Y.; Peng,~Q. Stable and 7.7 wt
\mciteBstWouldAddEndPuncttrue
\mciteSetBstMidEndSepPunct{\mcitedefaultmidpunct}
{\mcitedefaultendpunct}{\mcitedefaultseppunct}\relax
\EndOfBibitem
\bibitem[Singh \latin{et~al.}(2023)Singh, Shukla, and Chakraborty]{singh2023highly}
Singh,~M.; Shukla,~A.; Chakraborty,~B. Highly efficient hydrogen storage of a Sc decorated biphenylene monolayer near ambient temperature: ab initio simulations. \emph{Sustainable Energy \& Fuels} \textbf{2023}, \emph{7}, 996--1010\relax
\mciteBstWouldAddEndPuncttrue
\mciteSetBstMidEndSepPunct{\mcitedefaultmidpunct}
{\mcitedefaultendpunct}{\mcitedefaultseppunct}\relax
\EndOfBibitem
\bibitem[Liu \latin{et~al.}(2025)Liu, Chen, Liao, Zhang, and Laranjeira]{LIU2025105802}
Liu,~Z.; Chen,~X.; Liao,~Y.; Zhang,~L.; Laranjeira,~J.~A. First-principles insights of na-decorated B7N5 monolayer for advanced hydrogen storage. \emph{Surfaces and Interfaces} \textbf{2025}, \emph{58}, 105802\relax
\mciteBstWouldAddEndPuncttrue
\mciteSetBstMidEndSepPunct{\mcitedefaultmidpunct}
{\mcitedefaultendpunct}{\mcitedefaultseppunct}\relax
\EndOfBibitem
\bibitem[Duan \latin{et~al.}(2024)Duan, Shi, Yao, Liu, Diao, Lei, and Liu]{duan2024reversible}
Duan,~Z.; Shi,~S.; Yao,~C.; Liu,~X.; Diao,~K.; Lei,~D.; Liu,~Y. Reversible hydrogen storage with Na-modified Irida-Graphene: A density functional theory study. \emph{International Journal of Hydrogen Energy} \textbf{2024}, \emph{85}, 1--11\relax
\mciteBstWouldAddEndPuncttrue
\mciteSetBstMidEndSepPunct{\mcitedefaultmidpunct}
{\mcitedefaultendpunct}{\mcitedefaultseppunct}\relax
\EndOfBibitem
\bibitem[Wang \latin{et~al.}(2020)Wang, Xu, Deng, Wu, Meng, Huang, Bi, Yang, and Lu]{wang2020lithium}
Wang,~Y.; Xu,~G.; Deng,~S.; Wu,~Q.; Meng,~Z.; Huang,~X.; Bi,~L.; Yang,~Z.; Lu,~R. Lithium and sodium decorated graphdiyne as a candidate for hydrogen storage: First-principles and grand canonical Monte Carlo study. \emph{Applied Surface Science} \textbf{2020}, \emph{509}, 144855\relax
\mciteBstWouldAddEndPuncttrue
\mciteSetBstMidEndSepPunct{\mcitedefaultmidpunct}
{\mcitedefaultendpunct}{\mcitedefaultseppunct}\relax
\EndOfBibitem
\bibitem[Martins \latin{et~al.}(2024)Martins, Maia, Laranjeira, Fabris, Albuquerque, and Sambrano]{MARTINS202498}
Martins,~N.~F.; Maia,~A.~S.; Laranjeira,~J.~A.; Fabris,~G.~S.; Albuquerque,~A.~R.; Sambrano,~J.~R. Hydrogen storage on the lithium and sodium-decorated inorganic graphenylene. \emph{International Journal of Hydrogen Energy} \textbf{2024}, \emph{51}, 98--107\relax
\mciteBstWouldAddEndPuncttrue
\mciteSetBstMidEndSepPunct{\mcitedefaultmidpunct}
{\mcitedefaultendpunct}{\mcitedefaultseppunct}\relax
\EndOfBibitem
\bibitem[Djebablia \latin{et~al.}(2024)Djebablia, Abdullahi, Zanat, and Ersan]{djebablia2024metal}
Djebablia,~I.; Abdullahi,~Y.~Z.; Zanat,~K.; Ersan,~F. Metal-decorated boron phosphide (BP) biphenylene and graphenylene networks for ultrahigh hydrogen storage. \emph{International Journal of Hydrogen Energy} \textbf{2024}, \emph{66}, 33--39\relax
\mciteBstWouldAddEndPuncttrue
\mciteSetBstMidEndSepPunct{\mcitedefaultmidpunct}
{\mcitedefaultendpunct}{\mcitedefaultseppunct}\relax
\EndOfBibitem
\bibitem[Gong \latin{et~al.}(2024)Gong, Chen, Guo, Chen, Zhu, and Cheng]{gong2024me}
Gong,~Y.; Chen,~D.; Guo,~B.; Chen,~S.; Zhu,~Z.; Cheng,~M. Me--C8B5: A novel two-dimensional carbon boride as reversible hydrogen storage material with high capacity. \emph{International Journal of Hydrogen Energy} \textbf{2024}, \emph{82}, 384--397\relax
\mciteBstWouldAddEndPuncttrue
\mciteSetBstMidEndSepPunct{\mcitedefaultmidpunct}
{\mcitedefaultendpunct}{\mcitedefaultseppunct}\relax
\EndOfBibitem
\bibitem[Boubkri \latin{et~al.}(2024)Boubkri, Kassaoui, Razouk, Balli, and Mounkachi]{boubkri2024computational}
Boubkri,~M.; Kassaoui,~M.~E.; Razouk,~A.; Balli,~M.; Mounkachi,~O. Computational investigation of NLi4-cluster decorated phosphorene for reversible hydrogen storage. \emph{International Journal of Hydrogen Energy} \textbf{2024}, \emph{72}, 1--8\relax
\mciteBstWouldAddEndPuncttrue
\mciteSetBstMidEndSepPunct{\mcitedefaultmidpunct}
{\mcitedefaultendpunct}{\mcitedefaultseppunct}\relax
\EndOfBibitem
\bibitem[Guo \latin{et~al.}(2024)Guo, Chen, Chen, Song, Chen, Lin, and Cheng]{guo2024dtt}
Guo,~B.; Chen,~D.; Chen,~R.; Song,~C.; Chen,~Y.; Lin,~Q.; Cheng,~M. DTT graphene: A novel Dirac semimetal with ultrahigh reversible hydrogen storage capacity. \emph{International Journal of Hydrogen Energy} \textbf{2024}, \emph{58}, 987--999\relax
\mciteBstWouldAddEndPuncttrue
\mciteSetBstMidEndSepPunct{\mcitedefaultmidpunct}
{\mcitedefaultendpunct}{\mcitedefaultseppunct}\relax
\EndOfBibitem
\bibitem[Kumar \latin{et~al.}(2024)Kumar, Sharma, Pandey, and Tit]{KUMAR2024510}
Kumar,~N.; Sharma,~M.; Pandey,~R.; Tit,~N. Borophene/graphene heterostructure for effective hydrogen storage with facile dehydrogenation. \emph{International Journal of Hydrogen Energy} \textbf{2024}, \emph{70}, 510--521\relax
\mciteBstWouldAddEndPuncttrue
\mciteSetBstMidEndSepPunct{\mcitedefaultmidpunct}
{\mcitedefaultendpunct}{\mcitedefaultseppunct}\relax
\EndOfBibitem
\end{mcitethebibliography}

\end{document}